\documentclass[11pt,twocolumn]{article}
\usepackage{graphicx}
\usepackage[left=1.80cm, right=1.80cm, top=2.00cm, bottom=2.00cm]{geometry}
\usepackage{amsmath,bm}
\usepackage[colorlinks]{hyperref}
\usepackage[natbib=true, sorting=none, backend=biber, bibencoding=utf8, style=ieee]{biblatex}
\usepackage{eurosym}
\usepackage[dvipsnames]{xcolor}
\usepackage{subcaption}
\usepackage{booktabs} 
\usepackage{enumitem} 
\usepackage{accents}
\usepackage[capitalise]{cleveref}

\graphicspath{{figures/}{figures/example/}}

\crefname{relation}{Rel.}{Rels.}
\creflabelformat{relation}{(#2#1#3)}
\crefname{constraint}{Constr.}{Constrs.}
\creflabelformat{constraint}{(#2#1#3)}
\crefname{subequations}{Eqs.}{Eqs.}
\creflabelformat{subequations}{(#2#1#3)}

\newcommand{\ie}{\textit{i.e.} }

\newcommand{\ubar}[1]{\underaccent{\bar}{#1}}

\newcommand{\vpad}{\vspace{1mm}}
\newcommand{\hpad}{\hspace{15pt}}
\newcommand{\resultsin}[1]{\hspace{6pt} \bot  \hspace{6pt} #1}
\newcommand{\Forall}[1]{\hspace{10pt} \forall \,\, #1 }
\newcommand{\pdv}[2]{\frac{\partial #1}{\partial #2}}

\newcommand{\state}[1][i]{s_{#1,t}}
\newcommand{\capacity}{S_{i}}
\newcommand{\costfactor}{\gamma^\circ_{i,t}}
\newcommand{\capacityupper}{\bar{S}}
\newcommand{\capacitylower}{\ubar{S}}
\newcommand{\muuppernom}{\bar{\mu}^\text{nom}_{i}}
\newcommand{\mulowernom}{\ubar{\mu}^\text{nom}_{i}}

\newcommand{\kk}{k~\euro~}

\newcommand{\generation}{g_{s,t}}
\newcommand{\generationpotential}{\bar{g}_{s,t}}

\newcommand{\nodalgeneration}[1][n]{g_{#1,t}}
\newcommand{\capacitygeneration}{G_{s}}
\newcommand{\capacitygenerationupper}{\bar{G}_{s}}

\newcommand{\operationalpricegeneration}{o_{s}}
\newcommand{\capitalpricegeneration}{c_{s}}
\newcommand{\mulowergeneration}{\ubar{\mu}_{s,t}}
\newcommand{\muuppergeneration}{\bar{\mu}_{s,t}}
\newcommand{\muuppergenerationnom}{\bar{\mu}^\text{nom}_{s}}

\newcommand{\flow}{f_{\ell,t}}
\newcommand{\capacityflow}{F_{\ell}}

\newcommand{\operationalpriceflow}{o_\ell}
\newcommand{\capitalpriceflow}{c_{\ell}}
\newcommand{\mulowerflow}{\ubar{\mu}_{\ell,t}}
\newcommand{\muupperflow}{\bar{\mu}_{\ell,t}}

\newcommand{\storage}{g_{r,t}}
\newcommand{\storagedispatch}{\storage^\text{dis}}
\newcommand{\storagecharge}{\storage^\text{sto}}
\newcommand{\storagesoc}{\storage^\text{ene}}
\newcommand{\storageprevioussoc}{g_{r,t-1}^\text{ene}}

\newcommand{\efficiency}{\eta_{r}}
\newcommand{\efficiencydispatch}{\efficiency^\text{dis}}
\newcommand{\efficiencycharge}{\efficiency^\text{sto}}
\newcommand{\efficiencysoc}{\efficiency^\text{ene}}

\newcommand{\operationalpricestorage}{o_r}
\newcommand{\capitalpricestorage}{c_r}
\newcommand{\capacitystorage}{G_r}

\newcommand{\mulowerstoragedispatch}{\ubar{\mu}_{r,t}^\text{dis}}
\newcommand{\muupperstoragedispatch}{\bar{\mu}_{r,t}^\text{dis}}

\newcommand{\muupperstoragecharge}{\bar{\mu}_{r,t}^\text{sto}}

\newcommand{\muupperstoragesoc}{\bar{\mu}_{r,t}^\text{ene}}

\newcommand{\mustateofcharge}{\lambda^\text{ene}_{r,t}}

\newcommand{\lagrangian}{\mathcal{L}}
\newcommand{\lmp}[1][n]{\lambda_{#1,t}}
\newcommand{\averagelmp}{\bar{\lambda}_n}
\newcommand{\demand}[1][n]{d_{#1,t}}

\newcommand{\netconsumption}[1][n]{p^{-}_{#1,t}}
\newcommand{\netproduction}[1][n]{p^{+}_{#1,t}}

\newcommand{\incidence}[1][n]{K_{#1,\ell}}
\newcommand{\incidencegenerator}[1][n]{K_{#1,s}}
\newcommand{\incidencestorage}[1][n]{K_{#1,r}}
\newcommand{\incidenceasset}[1][n]{K_{#1,i}}
\newcommand{\ptdf}[1][n]{H_{\ell,#1}}
\newcommand{\cycle}{C_{\ell,c}}
\newcommand{\reactance}{x_\ell}
\newcommand{\cycleprice}{\lambda_{c,t}}

\newcommand{\emission}{e_{s}}
\newcommand{\emissionprice}{\mu_{\text{CO2}}}
\newcommand{\megawatthour}{MWh$_\text{el}$}
\newcommand{\totalcost}{\mathcal{T}}
\newcommand{\cost}[1][\circ]{\mathcal{C}^{#1}}
\newcommand{\payment}[1][n]{\mathcal{C}_{#1,t}}
\newcommand{\opex}{\mathcal{O}}
\newcommand{\opexgeneration}{\mathcal{O}^G}
\newcommand{\opexflow}{\mathcal{O}^F}
\newcommand{\opexstorage}{\mathcal{O}^E}
\newcommand{\capexgeneration}{\mathcal{I}^G}
\newcommand{\capexflow}{\mathcal{I}^F} 
\newcommand{\capexstorage}{\mathcal{I}^E}
\newcommand{\emissioncost}{\mathcal{E}}
\newcommand{\remainingcost}{\mathcal{R}}
\newcommand{\scarcitycost}{\remainingcost^\text{scarcity}}
\newcommand{\subsidycost}{\remainingcost^\text{subsidy}}

\newcommand{\allocatepeer}[1][m \rightarrow n]{A_{#1,t}}
\newcommand{\allocateflow}[1][n]{A_{\ell,#1,t}}

\newcommand{\allocatestate}[1][i, n]{A_{#1,t}}

\newcommand{\allocatecost}[1][n \rightarrow i]{\cost_{#1, t}}
\newcommand{\allocatecapex}[1][n \rightarrow s,t]{\mathcal{I}_{#1}}
\newcommand{\allocatecapexgeneration}[1][n \rightarrow s,t]{\capexgeneration_{#1}}

\newcommand{\allocateopex}[1][n \rightarrow s]{\opex_{#1,t}}

\newcommand{\allocatescarcitycost}[1][n \rightarrow i]{\scarcitycost_{#1,t}}

\begin{document}

\title{Tracing prices: A flow-based cost allocation for optimized power systems}
\author{Fabian Hofmann\\Frankfurt Institute for Advanced Studies (FIAS), 60438 Frankfurt, Germany\\
Email: hofmann@fias.uni-frankfurt.de}

\date{}

\twocolumn[
  \begin{@twocolumnfalse}

    \maketitle
    
    \begin{abstract}
        Power system models are a valuable and widely used tool to determine cost-minimal future operation and investment under political or ecological boundary conditions. Yet they are silent about the allocation of costs of single assets, as generators or transmission lines, to consumers in the network. Existing cost-allocation methods hardly suit large networks and do not take all relevant costs into account. This paper bridges this gap. Based on flow tracing, it introduces a peer-to-peer or more precisely an asset-to-consumer allocation of all costs in an optimized power system. The resulting cost allocation is both locally constrained and aligned with locational marginal prices in the optimum. The approach is applied and discussed using a future German scenario.
        \vtop{%
         \vskip0pt
          \hbox{%
          \includegraphics[width=\linewidth]{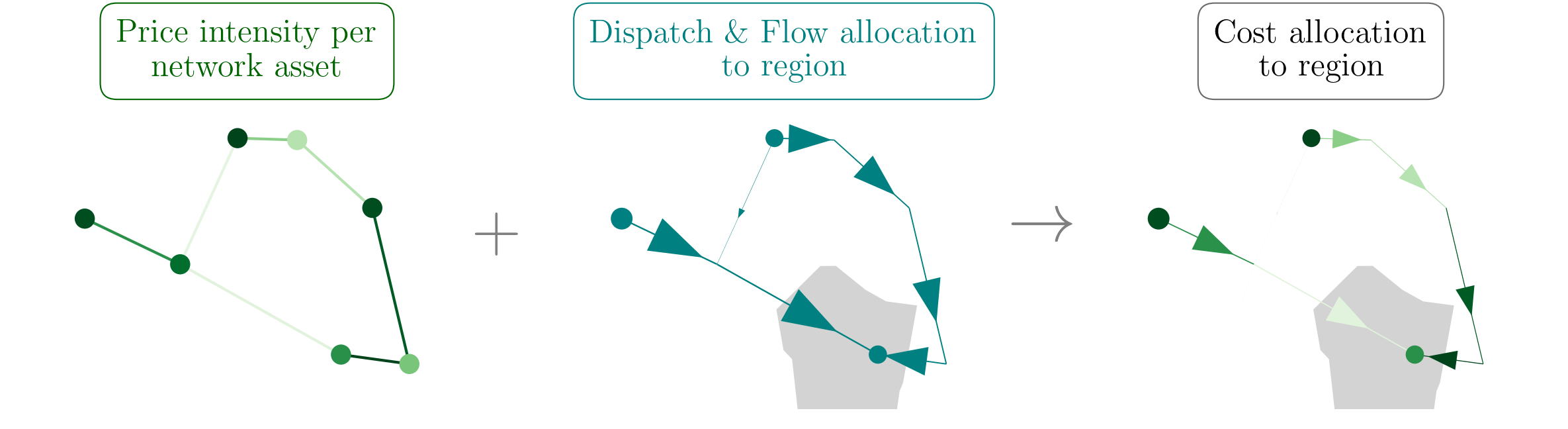}
          }%
        }
        
    \end{abstract}
 
\end{@twocolumnfalse}
]

\subsection*{Highlights}
\begin{itemize}
    \item In long-term equilibria a network asset recovers its variable and investment cost from its operation based revenue
    \item Flow tracing is used as a basis to allocate the operation of assets to consumers. 
    \item Allocated flows have to be reshaped such that the Kirchhoff Current Law and the Kirchhoff Voltage Law are respected. 
    \item Using operational and shadow prices from constraints, all costs are assigned in an P2P manner.
    \item The cost assignments are locally constraint and in alignment with the nodal pricing scheme of the optimum.    
    \item In a low-carbon scenario for Germany, regions with high renewable potentials profit from investments compensated by remote buses.   
\end{itemize}

\subsubsection*{Nomenclature}

\begin{table}[h]
    \centering
    \begin{tabular}{ll}
        $\lmp$ & Locational Market Price at bus $n$ and  \\ & time step $t$ in \euro/MW \vpad \\
        $\demand$ & Electric demand per bus $n$ \\ & time step $t$ in MW  \vpad \\
        $\state$ & Operational state of asset $i$, \\ & at time step $t$  in MW \vpad \\
        $o_i$ & Operational price of asset $i$ in \euro/MWh \vpad \\
        $c_i$ & Capital Price of asset $i$ in \euro/MW \vpad \\
        $\incidenceasset$ & Incidence values ($\ne$0  if $i$ is attached to $n$)  \vpad 
    \end{tabular}
\end{table}

\section{Introduction}

Today's power systems are subject to a deep and ongoing transformation. The shift from controllable to variable, weather-driven power generation as well as the constant improvement and innovation of technology require rigorous system planning and international cooperation \cite{pfenninger_energy_2014,schlachtberger_benefits_2017}. The core of the challenge manifests in the total costs of the system. Firstly, these should be as low as possible while meeting ecological and techno-economic standards. Secondly, they must be distributed in a fair and transparent manner among all agents in the power system. It is central to identify the drivers of costs and address them appropriately.  
In this respect, power system models are a valuable and widely used tool \cite{pfenninger_energy_2014,bazmi_sustainable_2011,pereira_generation_2017,brown_sectoral_2019}. Many studies for countries and regions throughout the world exist that lay out how renewable energy penetration can be expanded at minimum costs. Yet they largely remain silent how and on which grounds these costs are allocated among consumers. 

This paper fills this gap. In an optimized network, the characteristics of each time step are in detailed considered to allocate all system costs. We build on two fundamental concepts: first, the zero profit condition that states that, in the optimum, the revenue of each network asset, i.e., generator, transmission line etc., matches its operational and capital expenditures. Second, the flow tracing method, following Bialek’s Average Participation (AP) approach \cite{bialek_tracing_1996}, that allocates the use of network assets to consumers in a locally constrained fashion. 
Combining these two concepts allows for an transparent allocation of all operational (OPEX) and capital expenditures (CAPEX) to the consumers in the network. 
 
The literature discussed and applied the concept of flow-based cost allocation in a range of papers \cite{galiana_transmission_2003,shahidehpour_market_2002,meng_investigation_2007,schafer_allocation_2017,nikoukar_transmission_2012,arabali_pricing_2012,wu_locational_2005}. Shahidehpour et al. provide a profound insight into allocating congestion cost and transmission investments to market participants using different allocation techniques \cite{shahidehpour_market_2002}. Specifically, they set out that Generation Shift Factors, i.e. the marginal contribution of generators to a flow on line, allow to represent locational marginal prices (LMP) as a superposition of the LMP at the reference bus, the price for congestion, and a price for losses. The approach in \cite{meng_investigation_2007} expands this relation for contributions based on the AP scheme, which allows for an accurate estimation of the LMP, however it does not reflect the exact LMP of an optimized network. A similar approach is used in \cite{schafer_allocation_2017} that allocates electricity prices of a non-optimal power dispatch using flow tracing.

In this paper, we bring together the advantages of the studies discussed above. Our approach assures localized cost allocations while fully aligning payments to the nodal pricing scheme based on the LMP. It serves to facilitate transparency and cost-benefit analysis in network plannings such as the Ten Year Network Development Plan \cite{entso-e_completing_2020} or the German Netzentwicklungsplan (NEP) \cite{bundesnetzagentur_netzentwicklungsplan_2020}. Further, it provides a point of departure for usage-based transmission cost allocation.

The first section formulates the operation based revenue for different kind of assets (\cref{sec:zero_profit_rule}),  the AP allocation scheme and derived allocations from asset to consumer (\cref{sec:dispatch_allocations}), the cost allocation and the impact of additional constraints (\cref{sec:general_scheme,sec:design_constraints}), a numerical example (\cref{sec:numerical_example}). \Cref{sec:application_case} applies the cost allocation to a optimized German power system with a high share of renewable power generation and evaluates the allocated costs.

\section{Flow-based Cost Allocation}

\subsection{Operation based revenue}
\label{sec:zero_profit_rule}

\begin{figure}[h]
    \centering
    \includegraphics[width=.8\linewidth]{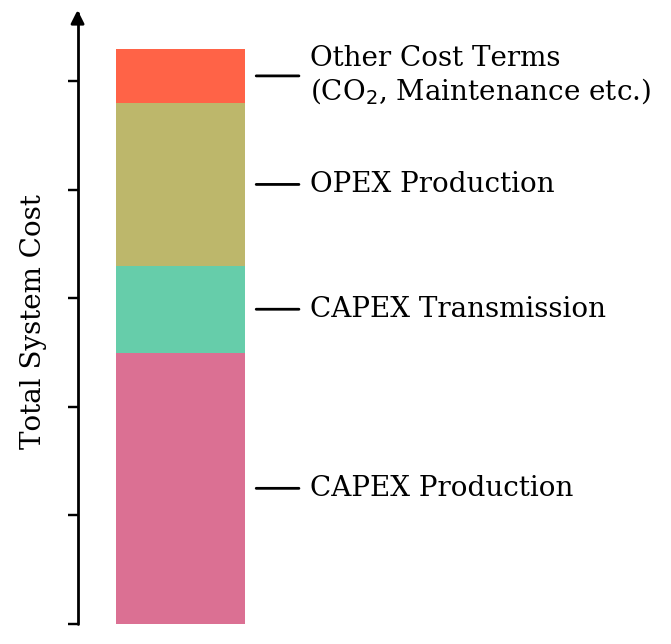}
    \caption{Schematic decomposition of the total system cost $\totalcost$ in a long-term investment model.}
    \label{fig:cost_decomposition}
\end{figure}
    
In long-term operation and investment planning models, the total costs $\totalcost$ of a power system is the sum of multiple cost terms $\cost$, as schematically depicted in \cref{fig:cost_decomposition}. Typically, these include operational expenditures for generators $\opexgeneration$, expenditures for emissions $\emissioncost$,  capital investments for the transmission system $\capexflow$ and so on, \ie
\begin{align}
\totalcost &= \sum_\circ \cost = \opexgeneration + \emissioncost +  \capexflow + ...
\label{eq:total_cost}
\end{align}

In turn, each of the terms $\cost$ consists of the costs associated to an asset $i$ in the system, 
\begin{align}
    \cost = \sum_{i} \cost_{i}
    \label{eq:asset_cost}
\end{align} 
where an ``asset'' describes any operating components of the network, such as a generator, line, energy storage etc. We refer to the set of all assets as $I$. 
In a long-term equilibrium of a power system with perfect competition and no further constraints, the zero-profit condition states that each cost term $\cost_{i}$ is recovered by the revenue that asset $i$ receives from the market \cite{steiner_peak_1957}. It builds on the fact that $\cost_i$ can be expressed as a cost-weighted sum of the operational state $s_{i,t}$ of asset $i$ and time $t$, \ie
\begin{align}
    \cost_{i} = \sum_t  \costfactor \, \state
    \label{eq:cost_decomposition}
\end{align}
where $\costfactor$ denotes a corresponding cost factor in \euro/MW. 
If $\cost_i$ describes the OPEX occasioned by asset $i$, the cost factor $\costfactor$ is simply given by the marginal operational price $o_i$. However, as we will show in the following, if it describes the CAPEX of asset $i$, $\costfactor$ is a composition of shadow prices $\mu_{i,t}$ given from the corresponding constraint at the cost-optimum. 

For a detailed demonstration, we derive  \cref{eq:cost_decomposition} for generators, transmission lines and storages separately. Therefore, let $o_i$ denote the operational price per MWh of asset $i$ and $c_i$ the capital price for one MW capacity expansion. \Cref{tab:cost_allocation_map} summarizes all derived relations. These can be inserted into \cref{eq:asset_cost,eq:cost_decomposition} for each single cost term.

\begin{table*}[t]
    \begin{center}
        \begin{tabular}{l|c|c|c|c|c}
        & $i$ & $\cost$ & $\cost_i$  & $\costfactor$ & $\state$  \\
        \toprule 
        OPEX Production & $s$ & $\opexgeneration$ & $\sum_{t} \operationalpricegeneration \, \generation$   & $\operationalpricegeneration$ & $\generation$ \\  
        OPEX Transmission  & $\ell$ & $\opexflow$ & $\sum_{t} \operationalpriceflow \, |\flow|  $ & $\operationalpriceflow$ & $|\flow|$ \\  
        OPEX Storage & $r$  & $\opexstorage$ & $\sum_{t} \operationalpricestorage \, \storagedispatch$ &  $\operationalpricestorage$ & $\storage$ \\
        \midrule
        Emission Cost & $s$ & $\emissioncost$ & $ \emissionprice \, \emission \, \generation$ & $\emissionprice \,\emission$ & $\generation$ \\
        \midrule   
        CAPEX Production & $s$ & $\capexgeneration$ & $ \capitalpricegeneration \capacitygeneration$ & $\muuppergeneration$ & $\generation$ \\
        CAPEX Transmission & $\ell$ & $\capexflow$ & $ \capitalpriceflow \capacityflow$ & $\left(\muupperflow - \mulowerflow \right)$ & $\flow$ \\
        CAPEX Storage & $r$ & $\capexstorage$ & $ \capitalpricestorage \capacitystorage$ & $ \muupperstoragedispatch - \mulowerstoragedispatch  + (\efficiencydispatch )^{-1} \mustateofcharge $ & $\storage$ \\
    \end{tabular}
    \end{center}
    \caption{Representation of different cost terms as a function of the operational state, \ie matching the form in \cref{eq:cost_decomposition}. These include OPEX \& CAPEX for production, transmission and storage assets in the network, as well as a cost term for the total Green House Gas (GHG) emissions. For storage units, an additional cost term adds to $\capexstorage$ (see \cref{sec:zero_profit_storage_units} for details).}
    \label{tab:cost_allocation_map}
\end{table*}

\subsubsection*{Generators}
\begin{subequations}
Let $S \subseteq I$ be the set of generators in the network, such that $\generation = s_{s,t}$ describes the power production of generator $s \in S$. The OPEX occasioned by generator $s$ is given by a cost-weighted sum of the production, thus 
\begin{align}
    \opexgeneration_s = \sum_t \operationalpricegeneration \, \generation 
    \label{eq:opexgeneration}
\end{align}
In case a fix price for emissions $\emissionprice$ in \euro\, per tonne-CO$_2$ equivalents, is assumed, a further the cost term per generator $s$,   
\begin{align}
 \emissioncost_s = \emissionprice \, \sum_t  \emission \, \generation
\end{align}
adds to $\totalcost$. Here, $\emission$ denotes the emission factor in tonne-CO$_2$ per \megawatthour\, of generator $s$. In contrast to OPEX and emission costs, the CAPEX of $s$ are not a function of the production $\generation$, but of the actual installed capacity $\capacitygeneration$ of generator $s$. In the optimization it limits the generation $\generation$ in the form of 
\begin{align}
\generation - \capacitygeneration  &\le 0 \resultsin{\muuppergeneration} \Forall{s,t} 
\label[constraint]{eq:upper_generation_capacity_constraint}
\end{align}
The constraint yields a shadow-price of $\muuppergeneration$, given by corresponding the Karush–Kuhn–Tucker (KKT) variable, in literature often denoted as the Quality of Supply \cite{schweppe_spot_1988}. It can be interpreted as the price per MW that \cref{eq:upper_generation_capacity_constraint} imposes to the system. If $\muuppergeneration$  is bigger than zero, the constraint is binding, which pushes investments in $\capacitygeneration$. As shown in \cite{brown_decreasing_2020} and in detail in \cref{sec:zero_profit_generation}, over the whole time span, the CAPEX for generator $s$ is recovered by the production $\generation$ times the shadow price $\muuppergeneration$, 
\begin{align}
 \capexgeneration_s = \capitalpricegeneration \capacitygeneration = \sum_t \muuppergeneration \,  \generation 
 \label{eq:no_profit_capex_generation}
\end{align}
This representation connects the CAPEX with the operational state of generator $s$, \ie matches the form in \cref{eq:cost_decomposition}.   
\end{subequations}

\subsubsection*{Transmission Lines}
\begin{subequations}
    
Let $L \subset I$ be the set of transmission lines in the system, these may include Alternating Current (AC) as well as Directed Current (DC) lines. Further let $\flow = s_{\ell,t}$ represent the power flow on line $\ell \in L$.  If the OPEX of the transmission system is taken into account in $\totalcost$ (these are often neglected in power system models), these may be approximated by $\opexflow_\ell = \sum_t o_\ell |\flow|$, that is, a cost weighted sum of the net flow on line $\ell$. Again this stands in contrast to the CAPEX which not a function of $\flow$ but of the transmission capacity $\capacityflow$. It limits the flow $\flow$ in both directions,
\begin{align}
\flow - \capacityflow &\le 0 \resultsin{\muupperflow} \Forall{\ell,t} 
\label[constraint]{eq:upper_flow_capacity_constraint} \\
- \flow - \capacityflow &\le 0 \resultsin{\mulowerflow} \Forall{\ell,t} 
\label[constraint]{eq:lower_flow_capacity_constraint}
\end{align}
At the cost-optimum, the two constraints yield the shadow prices $\muupperflow$ and $\mulowerflow$. Again, we use the relation that over the whole time span, the shadow prices weighted by the flow match the investment in line $\ell$ (for details see \cref{sec:zero_profit_flow}) 
\begin{align}
\capexflow_\ell = \capitalpriceflow \capacityflow = \sum_{t} \left( \muupperflow - \mulowerflow \right)  \flow 
\label{eq:no_profit_capex_flow}
\end{align}
The shadow prices $\muupperflow$ and $\mulowerflow$ can be seen as a measure for necessity of transmission investments at $\ell$ at time $t$. Hence, a non-zero values indicate that \cref{eq:upper_flow_capacity_constraint} or \eqref{eq:lower_flow_capacity_constraint} are bound and therefore that the congestion on line $\ell$ at time $t$ is imposing costs to the system. 

\end{subequations}

\subsubsection*{Storages}
\label{sec:storages}
\begin{subequations}

Let $R \subset I$ denote all storages in the system. In a simplified storage model, $\capacitystorage$ limits the storage dispatch $\storagedispatch$ and charging $\storagecharge$. Further it limits the maximal storage capacity $\storagesoc$ by a fix ratio $h_r$, denoting the maximum hours at full discharge. The storage $r$ dispatches power with efficiency $\efficiencydispatch$, charges power with efficiency $\efficiencycharge$ and preserves power from one time step $t$ to the next, $t+1$, with an efficiency of $\efficiencysoc$. In \cref{sec:zero_profit_storage_units} we formulate the mathematical details. The OPEX which adds to $\totalcost$ is given by 
\begin{align}
    \opexstorage = \sum_r o_r \storagedispatch 
\end{align}
Using the result of \cite{brown_decreasing_2020} the CAPEX can be related to the operation of a storage unit $r$ through
\begin{align}
    \notag
    \capexstorage =& \capitalpricestorage \, \capacitystorage \\
    \notag
    =& \sum_t \left(\muupperstoragedispatch - \mulowerstoragedispatch  + (\efficiencydispatch )^{-1} \mustateofcharge \right) \storagedispatch \\
    &- \sum_t \lmp \incidencestorage  \storagecharge \Forall{r} 
    \label{eq:no_profit_capex_storage}
\end{align}
where $\muupperstoragedispatch$ and $\mulowerstoragedispatch$ are the shadow prices of the upper and lower dispatch capacity bound and $\mustateofcharge$ is the shadow price of the energy balance constraint. Following the considerations in \cref{sec:zero_profit_storage_units} we restrict to the revenue from dispatched power, \ie the first term in \cref{eq:no_profit_capex_storage}, for the cost allocation.   

\end{subequations}

\subsection{Allocation of Power Dispatch}
\label{sec:dispatch_allocations}
\begin{subequations}
    
The fact that all asset related costs $\cost_i$ can be represented as a cost-weighted sum of the operation $\state$, prompts the question how $\state$ in turn is allocated to consumers. 

Dispatch and flow in a system can be considered as a superposition of individual contributions of nodes or assets. In order to artificially quantify these contribution, the literature provides various methods, named flow allocation schemes. Each of these follow a specific set of assumptions which lead to peer-to-peer allocations $A_{m \rightarrow n}$. That is a measure for the power produced at node $m$ and consumed at node $n$. 

In the following, we resort to one specific flow allocation scheme Average Participation (AP), also known as flow tracing \cite{bialek_tracing_1996}. The basic idea is to trace the power injection at bus $m$ through the network while following the real power flow on transmission lines and applying the principal of proportional sharing. At each bus, including the starting bus $m$, the traced flow might be mixed with incoming power flows from other buses. As soon as the power is absorbed or flowing out of a bus, the traced flow originating from $m$ splits in the same proportion as the total flow. This assumption leads to regionally confined allocation $A_{m \rightarrow n}$ based on a straightforward principal. A mathematical formulation of the AP scheme is documented in \cref{sec:net_ap}. For a detailed comparison with other schemes we refer to \cite{hofmann_techno-economic_2020}.

The peer-to-peer allocations $\allocatepeer$ fulfill some basic properties. On the one hand it allocates the all power productions at time $t$, \ie when summing over all receiving nodes $n$ the allocations yield the gross power generation $\nodalgeneration[m]$ of producing assets $i \in S \cup R$ (generators and storages) attached to $m$. Mathematically this translates to    
\begin{align}         
    \nodalgeneration[m] = 
    \sum_{i \in S \cup R} \incidenceasset[m] \, \state = \sum_n \allocatepeer
    \label{eq:nodalgeneration}
\end{align}
where $\incidenceasset[m]$ is 1 if asset $i$ is attached to bus $m$ and zero otherwise. On the other hand, when summing over all supplying nodes, the allocation  $\allocatepeer$ yields the gross nodal demand $\demand$ at node $n$ and time $t$, which leaves us with  
\begin{align}
    \demand = \sum_m \allocatepeer 
\end{align}
As the standard formulation implies, we assume that only net power production of $m$ is allocated to other buses, \ie $\allocatepeer[m \rightarrow m] = \min\left(\nodalgeneration[m], \demand[m]\right)$. Like this nodal generations which do not exceed the nodal demand are completely allocated to local consumers. 
The contribution of a single producing asset $i \in S \cup R$ to the nodal allocation $\allocatepeer$ is in proportion to the share $w_{i,t} = \incidenceasset[m] \state / \nodalgeneration[m]$ that asset $i$ contributes to the nodal generation $\nodalgeneration[m]$, leading to 
\begin{align}
    \allocatestate = w_{i,t}\, \allocatepeer \Forall{i \in S \cup R, n , t}
    \label{eq:allocate_production}
\end{align}
This relation states how much of the power produced by asset $i \in S \cup R$ is finally consumed by consumers at $n$. 

Yet, we didn't touch the allocation of transporting assets $i \in L$ to consumers. Note that the traced flow based in the AP scheme obeys the Kirchhoff Current Law but not the Kirchhoff Voltage Law, as already pointed out in \cite{galiana_transmission_2003}. However, as we show later a consideration of both laws is needed in order to align the allocated costs with the optimized locational marginal prices (LMP). To tackle this, let $\ptdf$ denote a linear mapping between the injection $\left(\nodalgeneration[m] - \demand[m]\right)$ and the flow $\flow$, such that  
\begin{align}
 \flow  = \sum_m \ptdf[m]\, \left(\nodalgeneration[m] - \demand[m]\right) \Forall{\ell \in L, t}
 \label{eq:power_flow_equation}  
\end{align}
Usually, $\ptdf$ is given by the Power Transfer Distribution Factors (PTDF) which indicate the changes in the flow on line $\ell$ for one unit (typically one MW) of net power production at bus $m$. For transport models or networks with High Voltage Directed Current (HVDC) lines, these can be set retrospectively using the formulation presented in \cite{hofmann_flow_2020}.
Inserting \cref{eq:nodalgeneration} into \cref{eq:power_flow_equation} and expanding the sum yields 
\begin{align}
    \allocateflow = \sum_m \ptdf[m] \left( \allocatepeer  - \delta_{n m} \demand \right) \Forall{\ell \in L, n, t}
 \label{eq:allocate_flow}
\end{align}
Complementary to \cref{eq:allocate_production}, this allocation indicates the power flow on line $\ell$ and time $t$ which is finally consumed by $\demand$.   

With \cref{eq:allocate_production,eq:allocate_flow} the allocation from all assets $i \in I$ to consumers in the network is derived. Naturally, the sum over all receiving nodes reproduces the operation $\state$ of asset $i$, 
\begin{align}
    \state = \sum_n \allocatestate \Forall{i, t}
    \label{eq:decompose_state}
\end{align} 

\end{subequations}

\subsection{Cost Allocation}
\label{sec:general_scheme}
\begin{subequations}
\label[subequations]{eq:general_scheme}

Using the presented relations, we are able to straightforwardly define the cost allocation. Therefore, we insert \cref{eq:decompose_state} in \cref{eq:cost_decomposition} and decompose the sum, which leads to 
\begin{align}
    \allocatecost = \costfactor \, \allocatestate
    \label{eq:cost_allocation}
\end{align}
By default, the full costs $\cost_i$ associated with asset $i$ are allocated, \ie $\cost_i = \sum_n \allocatecost$. Likewise all system costs are allocated, $\totalcost = \sum_{\circ, i, n, t} \allocatecost$. 

The cost allocation entails a further important property. In a cost-optimal setup, the LMP describes the change of costs for an incremental increase of electricity demand $\demand$ at node $n$ and time $t$ \cite{schweppe_spot_1988}. Mathematically this translates to the derivative of the total system cost $\totalcost$ with respect to the local demand $\demand$, $\lmp = \partial \totalcost /\partial \demand$. At the optimum this quantity is given by a shadow price of the nodal balance constraint (see \cref{sec:lmp} for details). Now, when summing over all assets $i$ and cost terms $\circ$, the cost allocation yields 
\begin{align}
    \payment = \sum_{\circ, i} \allocatecost  = \lmp\, \demand  \Forall{n,t}
    \label{eq:payment}
\end{align} 
which we refer to as the nodal payment or payment.  
This relation which is in detailed proven in \cref{sec:proof_equivalence} shows that the cost allocation is embedded in the optimized nodal pricing scheme, \ie the nodal payment exactly matches the payment determined by the LMP $\lmp$.

\end{subequations}



\subsection{Design Constraints}
\label{sec:design_constraints}
\begin{subequations}

Power system modelling does rarely follow a pure Greenfield approach with unlimited capacity expansion. Rather, today's models are setting various constraints defining socio-political or  technical requirements. However, this will alter the equality of total cost and total revenue. More precisely, each constraint $h_j$ (other then the nodal balance constraint) of the form 
\begin{align}
    h_j \left(\state, \capacity \right) < K
    \label{eq:design_constraint}
\end{align}
where $K$ is any non-zero constant and $\capacity$ denotes the nominal capacities of asset $i$, will  alter \cref{eq:cost_decomposition} to 
\begin{align}
    \cost_i - \remainingcost^\circ_i &= \sum_t \costfactor \, \state \\
    &= \sum_{n,t} \allocatecost
\end{align}
and result in a mismatch of $\remainingcost^\circ_i$ between the expenses and the revenue, therefore the cost allocation to asset $i$.
According to the nature of \cref{eq:design_constraint} and $\remainingcost_i$, it is either larger, equal or lower than $\cost$. Whereas \cref{eq:payment} still holds true, the total of payments do not return the system cost $\totalcost$ anymore. The total mismatch $\remainingcost$ is given by 
\begin{align}
    \remainingcost = \totalcost - \sum_{n,t} \payment
\end{align}
In the following we highlight two often used classes of designs constraints in the form of \cref{eq:design_constraint} and show how to consider them into the cost allocation.  

\end{subequations}

\subsubsection*{Capacity Expansion Limit}

\begin{subequations}
In real-world setups the capacity expansion of generators, lines or other assets are often limited. This might be due to land use restrictions or social acceptance considerations. 
When constraining the capacity $\capacity$ to an upper limit $\capacityupper$, in the form of 
\begin{align}
    \capacity - \capacityupper \le 0 \resultsin{\muuppernom} \Forall{i \in I}
\label{eq:capacityexpansionmaximum},
\end{align}
the zero profit condition alters as soon as the constraint becomes binding. Then, asset $i$ is payed an additional scarcity rent 
\begin{align}
    \scarcitycost_i = - \muuppernom \capacity \Forall{i \in I}
    \label{eq:scarcitycost}
\end{align}
This rent may account for different possible realms, as for example the increased market price in higher competed areas or additional costs for social or environmental compensation. To end this, the share in $\allocatecost$ which consumers pay for the scarcity rent can be recalculated by a correct weighting of the shadow price $\muuppergenerationnom$ with the capital price $c_i$, leading to 
\begin{align}
    \allocatescarcitycost = \dfrac{\muuppernom}{c_i + \muuppernom} \, \allocatecost \Forall{i}
\end{align}
\end{subequations}

\subsubsection*{Brownfield Constraints}

\begin{subequations}
In order to take already built infrastructure into account, the capacity $\capacity$ may be constrained by a minimum required capacity $\capacitylower_i$. This introduces a constraint of the form 
\begin{align}
    \capacitylower - \capacity  \le 0 \resultsin{\mulowernom} \hpad \forall{i \in I}
\label{eq:capacityexpansionminimum}
\end{align}
Again, such a setup alters the zero profit condition of asset $i$, as soon as the constraint becomes binding. 
In that case, asset $i$ does not collect enough revenue from $\allocatecost$ in order to compensate the CAPEX. The difference, given by 
\begin{align}
    \subsidycost_i = \mulowernom \capacity
    \Forall{i}
\end{align}
has to be subsidized by governments or communities or is simply ignored when investments are amortized. Note, it is rather futile wanting to allocate these cost to consumers as assets may not gain any revenue for their operational state, \ie where $\cost_i = \subsidycost_i $. 
\end{subequations}

\subsection{Numerical Example}
\label{sec:numerical_example}
\begin{figure*}[t]
    \centering
    \includegraphics[width=\linewidth]{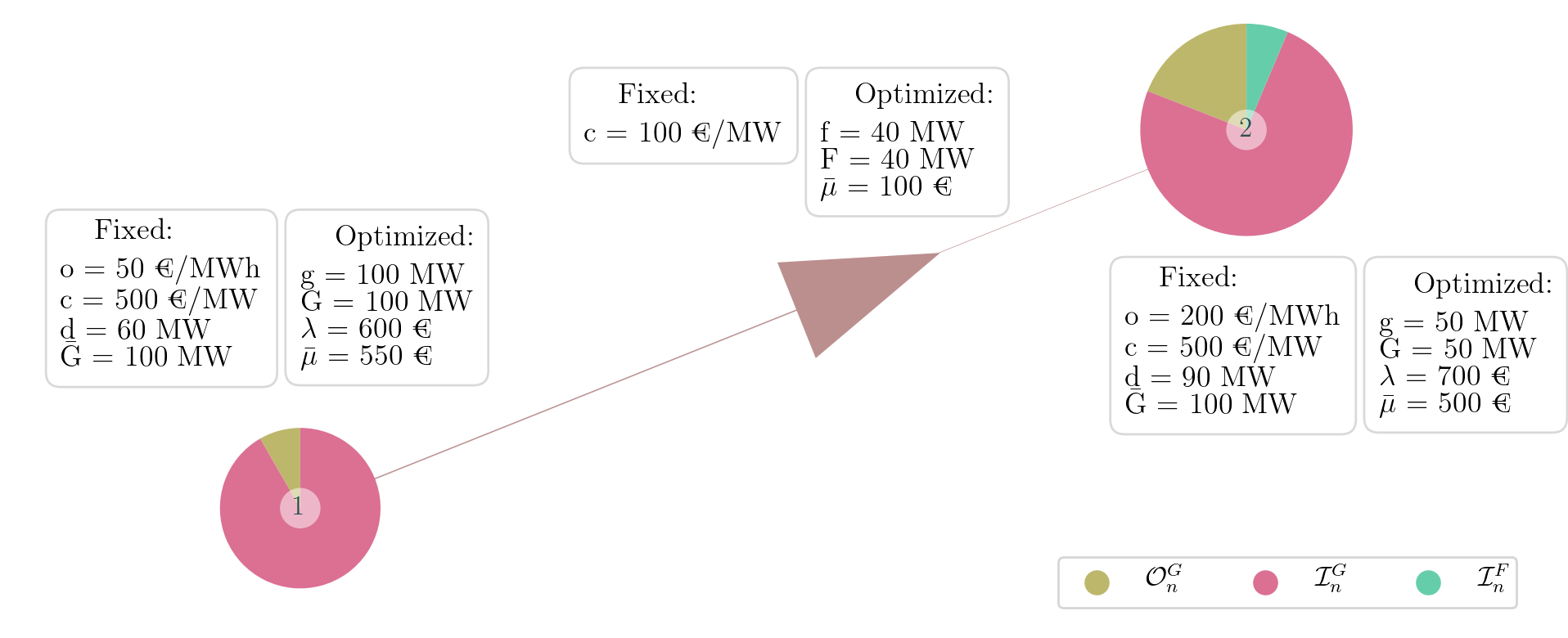}
    \caption{Illustrative example of a 2~bus network with one optimized time step. Fixed prices and constraining values are given in the left box for each bus and the transmission line. Optimized values are given in the right boxes. Generator~1 at bus~1 has a cheaper operational price $o$, capital prices are the same for both. As both generator capacities are constraint to 100~MW, the optimization also deploys the generator at bus~2. }
    \label{fig:example_network}
\end{figure*}
\begin{figure*}[h!]
    \begin{subfigure}[c]{.495\linewidth}
    \includegraphics[width=\linewidth]{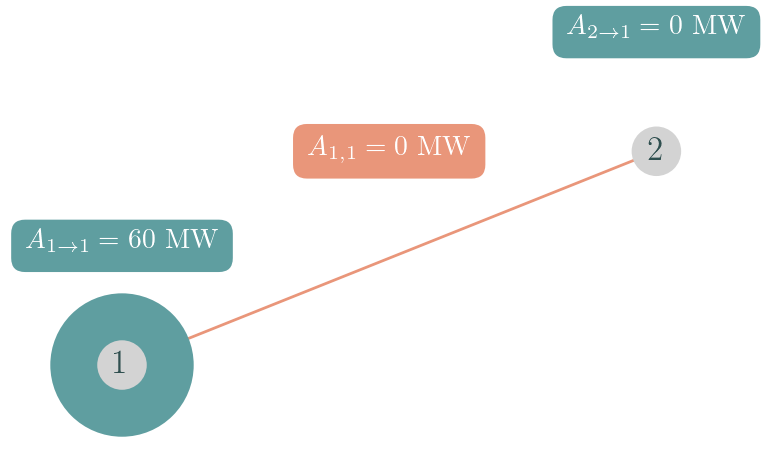}
    \vspace{-40pt}
    \subcaption{}
    \label{fig:example_allocation_bus1}
    \end{subfigure}
    \begin{subfigure}[c]{.495\linewidth}
    \includegraphics[width=\linewidth]{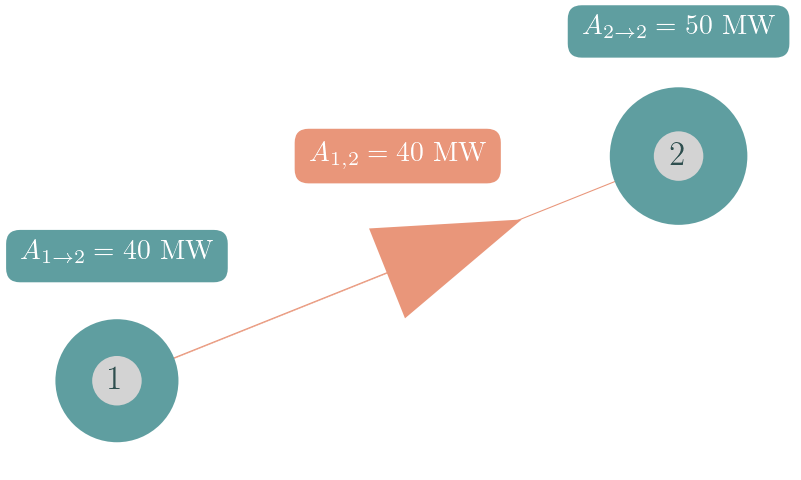}
    \vspace{-40pt}
    \subcaption{}
    \label{fig:example_allocation_bus2}
    \end{subfigure}
    \caption{Power allocations $A_{i,n}$ for the example network in \cref{fig:example_network} using Average Participation. The 60~MW consumption at bus~1 (a) are totally supplied by the local generator. In contrast consumers at bus~2 (b) retrieve 50~MW from the local generator, the remaining 40~MW are imported from generator~1 which induces a flow on the transmission line.}
    \label{fig:example_allocation}
\end{figure*}

\begin{figure}[h]
    \centering
    \includegraphics[width=\linewidth]{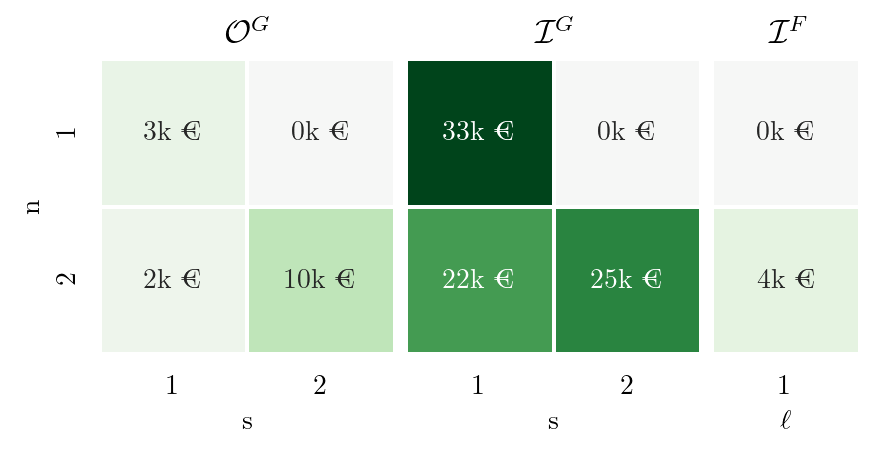}
    \caption{Full P2P cost allocation $\allocatecost$ for the optimized example network in \cref{fig:example_network}. Consumers compensate OPEX and CAPEX of the generators they retrieve from (compare with \cref{fig:example_allocation}). As bus~1 is totally self-supplying, all it payment is assigned to the local generator.  As bus~2 imports power from bus~1 and thus induces a flow on line~1, it not only compensates local expenditures but also OPEX and CAPEX of generator~1 and CAPEX of the transmission line.}
    \label{fig:example_payoff}
\end{figure}

After presenting the basics of the cost allocation, we present its behavior by means of a small numerical example.  
Consider a two bus system, depicted in \cref{fig:example_network}, with one transmission line and one generator per bus. Generator~1 (at bus~1) has a cheap operational price of 50 \euro/\megawatthour, generator~2 (at bus~2) has a expensive operational price of 200~\euro/\megawatthour. For both, capital investments amount 500~\euro/MW and the maximal capacity is limited to $\capacitygenerationupper$~=~100~MW. The transmission line has a capital price of 100~\euro/MW and no upper capacity limit. With a demand of 60~MW at bus~1 and 90~MW at bus~2, the optimization expands the cheaper generator at bus~1 to its full limit of 100~MW. The 40~MW excess power, not consumed at bus~1, flow to bus~2 where the generator is built with only 50~MW. \\
\Cref{fig:example_allocation} shows the allocation $A_{i,n}$ for bus~1 and bus~2 separately. The ``sum'' of the two figures give to the actual dispatch and flow. The resulting P2P payments are given in \cref{fig:example_payoff}.  

The left graph \cref{fig:example_allocation_bus1} shows that $d_1$ is totally supplied by the local production. Consequently consumers at bus~1 pay 3k~\euro~OPEX to generator~1, which is the operational price of 50~\euro\, times the retrieved power of 60~MW. Further they pay 33\kk of its CAPEX. Note that 3\kk of these account for the scarcity imposed buy the upper expansion limit $\bar{G}_1$. The rest makes up 60\% of the total CAPEX of generator~1, exactly the share of power allocated to $d_1$. Consumers at bus~1 don't pay any transmission CAPEX as no flow is assigned. \\
The right graph \cref{fig:example_allocation_bus2} shows the power allocations to $d_2$. We see that 50~MW are self-supplied whereas the remaining 40~MW are imported from generator~1. Thus, consumers pay for the local OPEX and CAPEX as well as the corresponding proportion of to generator~1 and the transmission line. As the capacity at generator~2 does not hit the expansion limit $\bar{G}_2$, no scarcity rent is assigned to it. The allocated $\allocatecapexgeneration[2\rightarrow 2]$  compensates  the full investment of generator~2. In contrast, 2\kk of the 22\kk which are allocated to investment in generator~1 are associated with the scarcity rent of generator~1. The payed congestion revenue of 4\kk is exactly the CAPEX of the transmission line.   

The sum of all values in the payoff matrix in \cref{fig:example_payoff} yield $\totalcost - \scarcitycost$, the total system cost minus the scarcity rent (which in turn is negative).
The sum of a column  yields the total revenue per the asset $i$. These values match their overall spending plus the cost of scarcity. The sum a row returns the nodal payments $\mathcal{C}_n = \lambda_n \, d_n$. For example the sum of payments of consumers at bus~1 is 36\kk. This is exactly the electricity price of 600~\euro/MW times the consumption of 60~MW. \\

The fact that OPEX and CAPEX allocations are proportional to each other results from optimizing one time step only. This coherence breaks for larger optimization problems with multiple time steps. Then CAPEX allocation takes effect only when one or more of the capacity constraints  \cref{eq:upper_generation_capacity_constraint,eq:lower_flow_capacity_constraint,eq:upper_flow_capacity_constraint} become binding.  \\

\section{Application Case}
\label{sec:application_case}
\begin{figure*}[t]
    \centering
    \includegraphics[width=\linewidth]{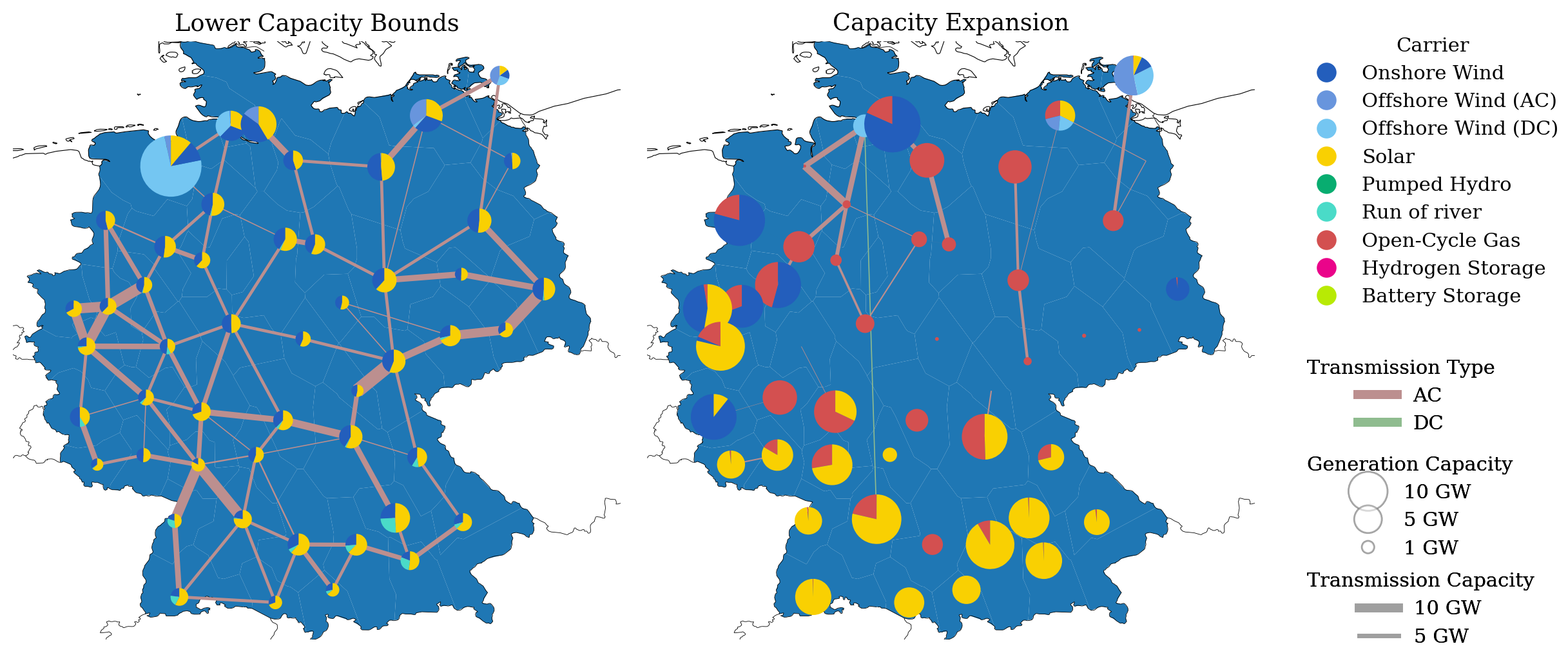}
    \caption{Brownfield optimization of the German power system. The left side shows existent renewable capacities, matching the total capacity for the year 2017, which serve as lower capacity limits for the optimization. The right side shows the capacity expansion of renewable resources as well as installation of backup gas power plants. The effective CO$_2$ price is set to 120 \euro\, per tonne CO$_2$ emission.}
    \label{fig:network}
\end{figure*}
We now showcase the behavior of the cost allocation in a more complex system and apply it to an cost-optimized German power system model with 50 nodes and one year time span with hourly resolution. The model builds up on the PyPSA-EUR workflow \cite{horsch_jonas_pypsa-eur_2020} with technical details and assumptions reported in \cite{horsch_pypsa-eur_2018}. 

We follow a brownfield approach where transmission lines can be expanded starting from today's capacity values, originally retrieved from the ENTSO-E Transmission System Map \cite{entso-e_entso-e_nodate}. Pre-installed wind and solar generation capacity totals of the year 2017 were distributed in proportion to the average power potential at each site excluding those with an average capacity factor of 10\%. Further, wind and solar capacity expansion are limited by land use restriction. These consider agriculture, urban, forested and protected areas based on the CORINE and NATURA2000 database \cite{corine2012,natura2000}. Pumped Hydro Storages (PHS) and Run-of-River power plants are fixed to today's capacities with no more expansion allowed. Additionally, unlimited expansion of batteries and H$_{2}$-storages and Open-Cycle Gas Turbines (OCGT) are allowed at each node. 
\begin{figure}
    \centering
    \includegraphics[width=\linewidth]{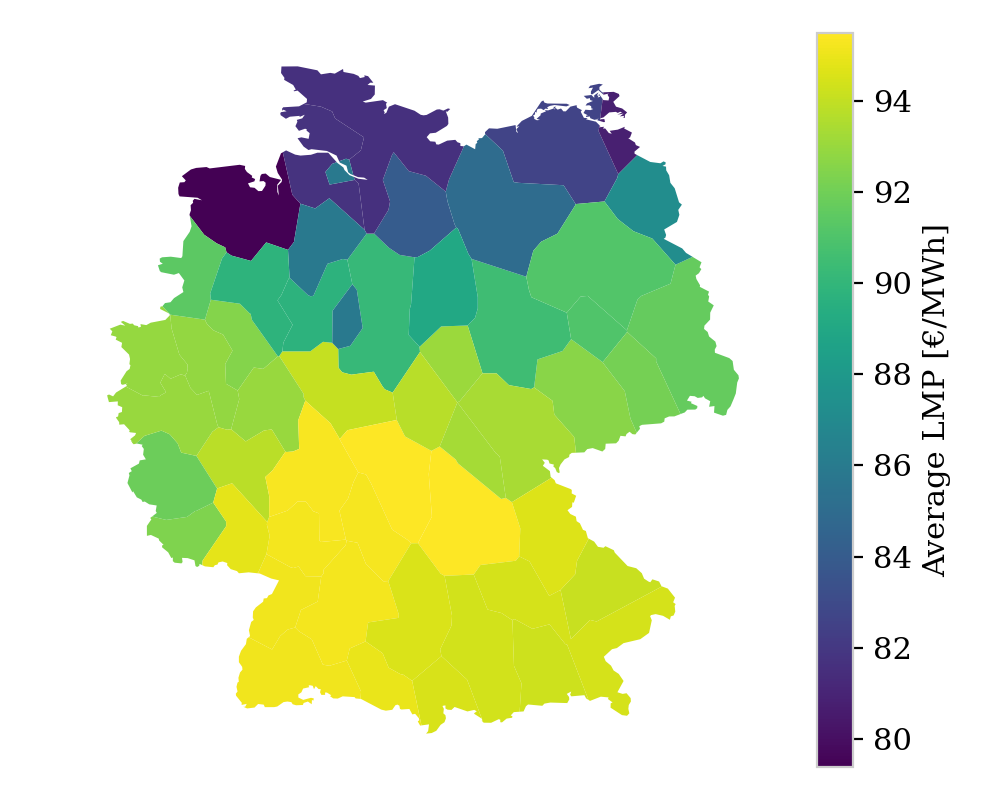}
    \caption{Load-weighted average electricity price $\averagelmp$ per region in the optimized German power system. Regions in the middle and south of Germany have high prices whereas electricity in the North with a strong wind, transmission and OCGT infrastructure is cheaper.}
    \label{fig:average_price}
\end{figure}
We impose an carbon price of 120 \euro\, per tonne-CO$_{2}$ which, for OCGT, adds an effective price of 55 \euro/\megawatthour (assuming a gross emission of 180~kg/MWh and an efficiency of 39\%). All cost assumptions on operational costs $o_i$ and annualized capital cost $c_i$ are summarized in \cref{tab:cost_assumptions}. 

The optimized network is shown in  \cref{fig:network}. On the left we find the lower capacity bounds for renewable generators and transmission infrastructure, on the right the optimized capacity expansion for generation, storage and transmission. Solar capacities are expanded in the south, onshore and offshore wind in the upper north and most west. Open-Cycle Gas Turbines (OCGT) are build within the broad middle of the network. Transmission lines are amplified in along the north-south axis, including one large DC link, associated with the German S\"ud-Link, leading from the coastal region to the southwest. 
The total annualized cost of the power system roughly sums up to 42 billion \euro.

\Cref{fig:average_price} displays the load-weighted average electricity price $\averagelmp$ per region, defined by 
\begin{align}
    \averagelmp = \dfrac{\sum_{t} \lmp \demand}{\sum_t \demand}    
\end{align}
We observe a relatively strong gradient from south (at roughly 92 \euro/MWh) to north (80 \euro/MWh). Regions with little pre-installed capacity and capacity expansion, especially with respect to renewable generation, tend to have higher prices. The node with the lowest LMP in the upper northwest, stands out through high pre-installed offshore capacities. \\
 
\begin{figure}
    \centering
    \includegraphics[width=0.85\linewidth]{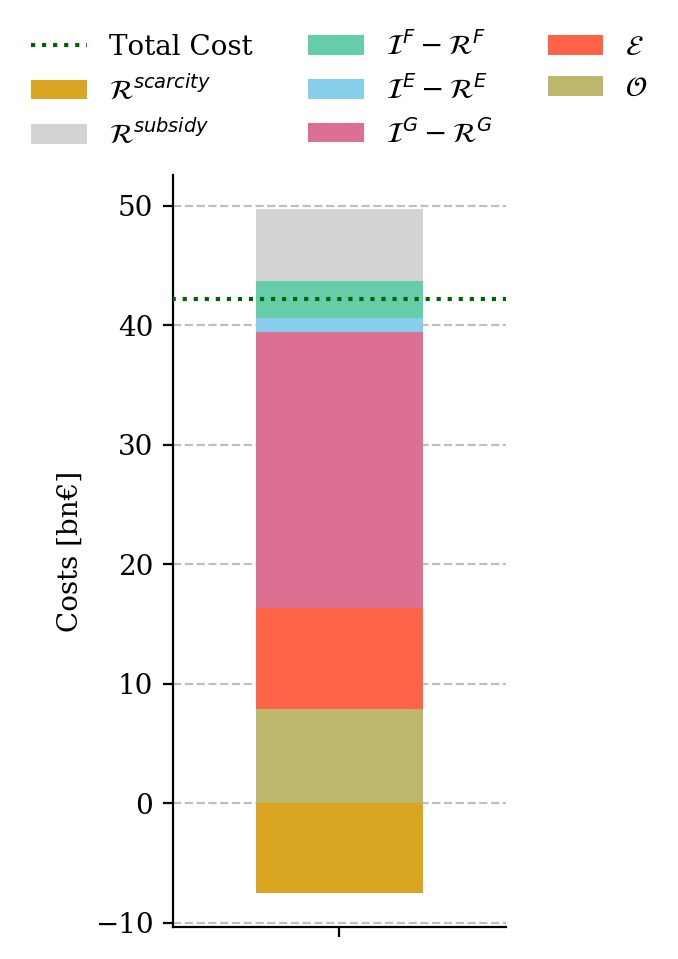}
    \caption{Total allocated payments of the system. }
    \label{fig:total_cost}
\end{figure}

In \cref{fig:total_cost} we show the total of all allocated costs. We split the CAPEX allocation into $\cost - \remainingcost^\circ$ and $\remainingcost^\circ$. The difference $\remainingcost^\circ$ consists of scarcity rents $\scarcitycost$ and subsidies $\subsidycost$. Note that the sum of all contributions in \cref{fig:total_cost} equals the total cost $\totalcost$.  
In the following, we address each of the displayed cost terms and its corresponding allocations separately.   

The largest proportion of the payments is associated with CAPEX for generators, transmission system and storage units in decreasing order. 
Taking production technologies into account, we observe a fundamental difference between the controllable OCGT  and the variable renewable resources: As shown in detail in \cref{fig:capex_duration_curve}, more than half of all investments in OCGT is determined in one specific hour. At this time (morning, end of February), the system hits the highest mismatch between low renewable power potentials and high demands. The necessity for backup generators manifests in high allocation of CAPEX to OCGT and consequently high LMP. With a few exceptions in the South West and East, all consumers receive power from OCGT at this time, thus all pay high amounts for the needed backup infrastructure (operational state of the system is in detailed shown in \cref{fig:operation_high_expenditure}). Note that this is the most extreme event, which ensures backup infrastructure for other inferior extreme events. The total CAPEX allocation for OCGT infrastructure is depicted in \cref{fig:ocgt_capex}. This  correlates with our findings of the extreme event. \\
Contrary to this, CAPEX for renewable infrastructure are allocated evenly throughout several thousands of hours. As for onshore and offshore wind farms, the produced power deeply penetrates the network, see \cref{fig:power_mix}, thus it is not only local, but also remote consumers which cover the CAPEX, see \cref{fig:capex_price}. This in turn benefits consumers, which profit from cheap operational prices of local wind farms. This explains why these regions end up with a low average LMP.\\

Together with the emission cost $\emissioncost$, the total OPEX $\opex$ amount around 16 billion \euro. As to expect, 99.97\% are dedicated to OCGT alone, as these have by far the highest operational price. For a detailed regional distribution of payments per MWh and the resulting revenues for generator, see \cref{fig:opex_price}. Since during ordinary demand peaks, it is rather local OCGT generators which serve as backup generators, thus the OPEX allocation of OCGT clearly differs from the CAPEX allocation. In the average power mix per region, see \cref{fig:power_mix}, we observe that regions with strong OCGT capacities also  have high shares of OCGT power consumption. The average operational price for renewable generators is extremely low (0.2 \euro/MWh), thus they play an inferior role in the OPEX allocation. \\

\begin{figure*}
    \centering
    \begin{subfigure}[c]{.6\linewidth}
    \includegraphics[width=\linewidth]{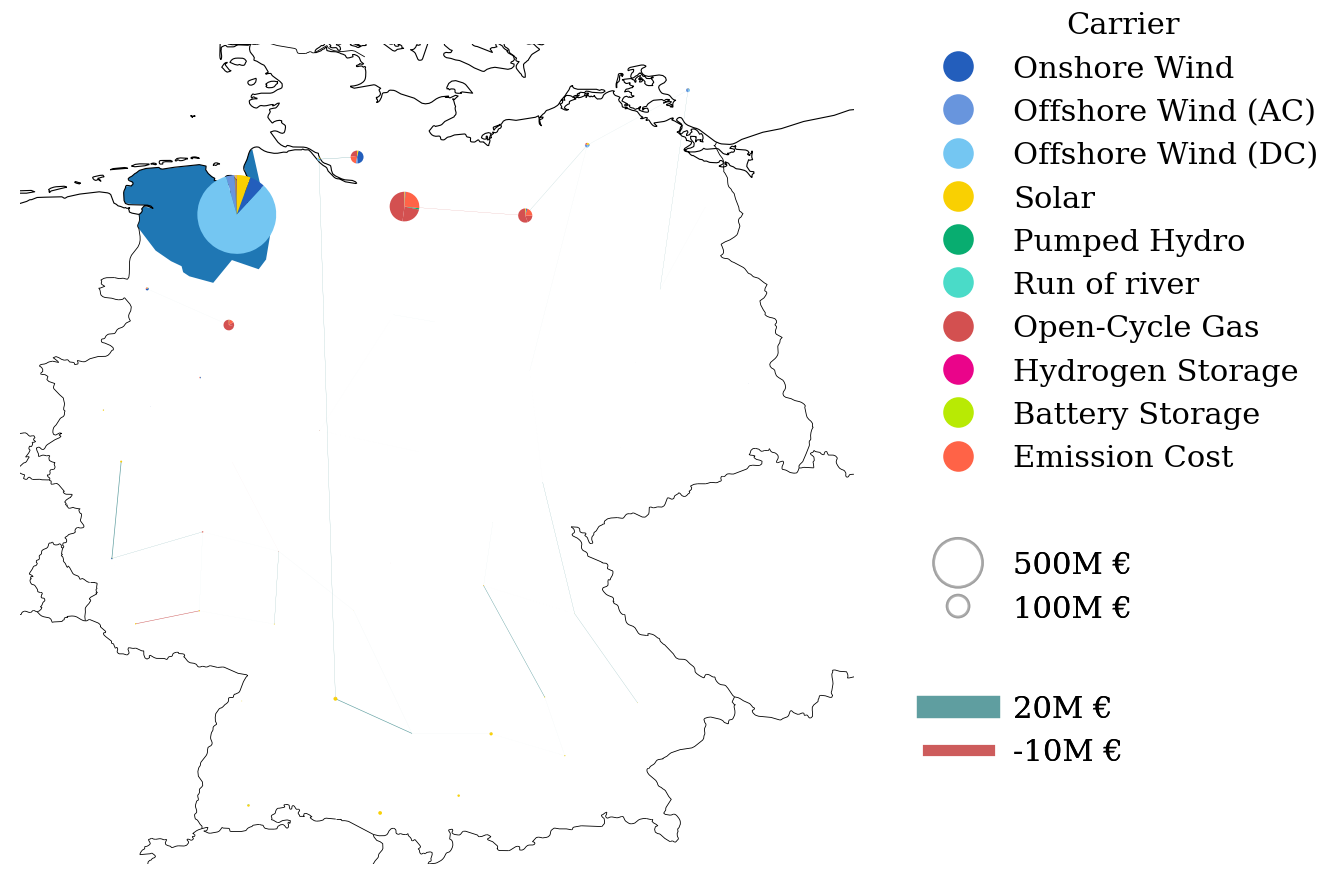}
    \end{subfigure}
    \hspace{-.2151\linewidth}
    \begin{subfigure}[c]{.6\linewidth}
    \includegraphics[width=\linewidth]{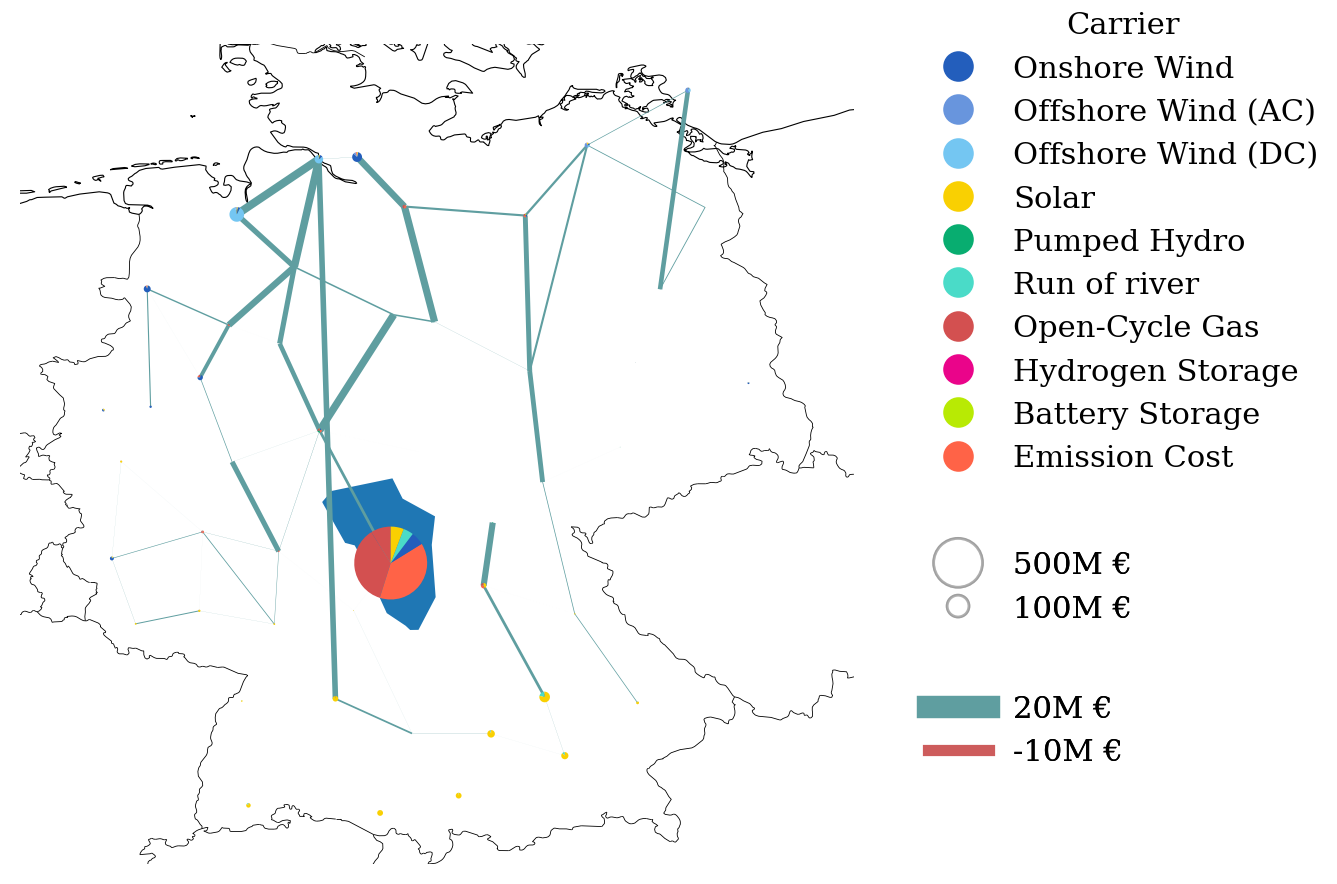}
    \end{subfigure}
    \caption{Comparison of payments of the region with the \textbf{lowest LMP (left)} and the region with the \textbf{highest LMP (right)}. The region is colored in dark blue. The circles indicate to which bus and technology OPEX and CAPEX are assigned. The thickness of the lines is proportional to dedicated payments.}
    \label{fig:direct-allocation}
\end{figure*}

\begin{figure}
    \centering
    \includegraphics[width=\linewidth]{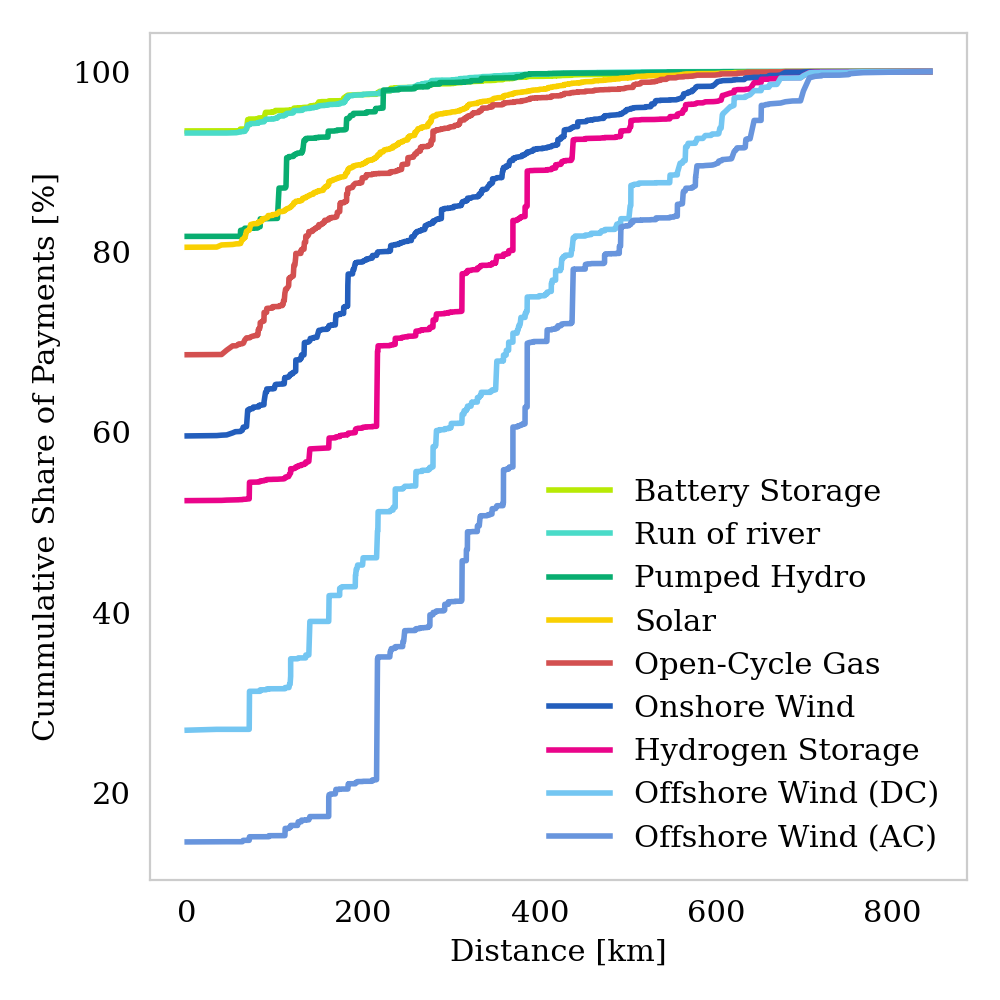}
    \caption{Average distance between payer and receiver for different technologies and shares of the total production.}
    \label{fig:locality}
\end{figure}

The negative segment in \cref{fig:total_cost} is associated with scarcity costs $\scarcitycost$, caused by land use constraints for renewable resources and the transmission expansion limit. These sum up to approximately 7.5bn~\euro. Note again, that according to \cref{eq:scarcitycost} this term is negative and is part of the allocated CAPEX payments. It translates to the cost that consumers pay ``too much'' for assets limited in their capacity expansion. In the real world this money would be spent for augmented land costs or civic participation in the dedicated areas. In \cref{fig:scarcity_price} we give a detailed insight of how the scarcity rent manifests in the average cost per consumed MWh. The scarcity for wind and solar is relatively low and contributes roughly 2~\euro/MWh to the load-weighted average price $\averagelmp$. It primarily affects regions in close vicinity to area with the highest renewable power potential. \\
Remarkably, the scarcity rent per MWh for run-of-river power plants amounts up to 16~\euro. This impact comes from the steady power potential from run-off water and the strong limitation of capacity expansion. However, as these power plants in particular are already amortized, the scarcity rent should be reconsidered and removed from a final cost allocation. \\
A high influence on the price comes from the scarcity of transmission expansion. Right beside regions with high wind infrastructure, it occasions average costs higher than MWh 4~\euro/MWh (maximal 8\euro/MWh), see \cref{fig:branch_scarcity_price}. As to expect, the constraint mainly suppresses transmission expansion along the north-south axis.   \\
The last cost term $\subsidycost$ in \cref{fig:total_cost}, is caused by lower capacity constraints for pre-existing assets. These violate the optimal design and are not recovered by the revenue. Most of these ``non-allocatable'' costs account for Pumped Hydro Storage, onshore wind, offshore wind and the transmission system, see \cref{fig:subsidy} for further details. Again, as most of these generators are amortized (PHS, transmission system), the costs should be reconsidered in a final cost allocation.

\Cref{fig:direct-allocation} compares the P2P cost assignments of the region with the lowest average LMP (left side) against the one with the highest LMP (right side). The low-price region in the north-west is fairly independent of investments in the transmission system. It profits from local offshore wind farms which are partly payed by subsidies, see \cref{fig:subsidy}. Only a small share of the payments is allocated to remote OCGTs. In contrast, the high-priced region is highly dependent on local OCGT and the transmission system, which causes high allocated OPEX, emission cost and transmission CAPEX. Its payments to  onshore and offshore wind infrastructure are low despite a third of its supply comes from wind power. Hence, the wind power supply at this region is not restricted by exhausted wind power resources but by bottlenecks in the transmission system.  

Finally, \cref{fig:locality} draws the cumulative share of P2P cost assignments of a function of the distance between producer and consumer. The data is shown for all technologies separately. The later the curve reaches 100\% the deeper the price of a technologies penetrates into the network. Offshore wind has the strongest price influence to remote buses. Only 10-30\% of its expenses are compensated by local consumers and the rest is assigned to remote buses, especially those with high demand (compare \cref{fig:average_demand}). The contrast between Hydrogen Storage and Battery sticks out. Whereas Battery costs are hardly assigned to other buses, almost 50\% of the Hydrogen Storage costs are payed by remote buses. It underlines the fundamental functionality of the two technologies. Hydrogen storage are located at buses with high wind generation and balance out their long-term excess and deficit energy. The dispatched power follows a similar way through the network as the wind power. The battery on the other hand pairs with local solar production and its locally constraint dispatch and flow.     

\section{Limitations}
The presented cost allocation is based on the linear power flow approximation. Yet, the method is equally applicable to a system with an Optimal Power Flow (OPF), \ie full AC power flows. Yet for this case, the AP scheme is not the correct choice as the misalignment from the Kirchhoff Voltage Law cannot not be corrected subsequently. Rather, the Z-Bus flow allocation presented in \cite{conejo_z-bus_2007} might suit better as it naturally respects both circuit laws. Allocating on the basis of the full power flow introduces an additional cost term $\remainingcost^\text{Loss}$ accounting for the transmission loss which is compensated by the consumers and mirrors in the LMP. 

The used optimization does not take security constrains of the transmission system into account. These may be incorporated following the approach in \cite{nikoukar_transmission_2012}. 

We restricted the application to long-term investment models with perfect foresight. However, the cost allocation can as well be applied to short-term planning models with fixed capacities. The revenue from the capacity limits then builds the basis for amortization and future investments.

The optimization assumes a fix demand time series. As shown in \cref{fig:capex_duration_curve} this leads to high if not unrealistic LMP. Introducing a value of loss load as proposed in \cite{schroder_value_2015}  would screen away these and lead to more evenly distributed allocations.

\section{Conclusion}

A new cost-allocation scheme based on peer-to-peer dispatch allocations from assets to consumers was presented. Within a long-term equilibrium OPEX and CAPEX of each asset are payed back by the operation based revenue. Using flow allocation, we are able to allocate the operation and therefore the assigned costs of assets to consumers in the network. For three typical classes of assets, namely generators, transmission lines and storage units, we showed how operational prices and shadow prices must be weighted with the dispatch allocation in order to allocate all system costs. Further we highlighted the impact of minimum capacity requirements and maximum installation potentials. These alter the revenues per asset and therefore the cost allocations. For lower capacity requirements, assets may not recover all the expenses from the revenue. In this case the cost difference has to be subsidized or simply be ignored, in case the asset is already amortized. Contrary, upper capacity expansion limits lead to an additional charge for consumers which have to compensate for an additional scarcity rent of the assets.  
Applied to a optimized German power system with an imposed price of 120 \euro\, per tonne CO$_2$ equivalent. The cost allocation shows low electricity prices for consumers in a renewable German system are achieved on a transparent basis. The cost allocation shows how buses remote from wind farms pay higher prices due to increased reliance in transmission and backup capacity. On the other hand, buses with high renewable installation spend most payments to local assets.

\subsubsection*{Reproducibility and Expansion}

All figures and data points can be reproduced by using the \textit{python} workflow in \cite{hofmann_pypsa-costallocation_2020}. The automated workflow allows for higher spatial resolution of the network (scalable up to a the full ENTSO-E Transmission System Map) and optionally taking the total European power system into account.  

\subsection*{Funding}
This research was funded by the by the German Federal Ministry for Economics Affairs and Energy (BMWi) in
the frame of the NetAllok project (grant number 03ET4046A) \cite{bundesministerium_fur_wirtschaft_und_energie_verbundvorhaben_nodate}. 

\subsubsection*{Acknowledgement}

In particular, I thank Tom Brown for very fruitful discussions. I am very grateful to Alexander Zerrahn for reviewing and helping out with important parts. Further, I want to thank Alexander Kies and Markus Schlott who steadily helped with creative thoughts.


\clearpage
\appendix

\section{Network Optimization}

\renewcommand\theequation{\thesection.\arabic{equation}}
\setcounter{equation}{0}

\renewcommand\thefigure{\thesection.\arabic{figure}}    
\setcounter{figure}{0}    

\subsection{LMP from Optimization}
\label{sec:lmp}
The nodal balance constraint ensures that the amount of power that flows into a bus equals the power that flows out of a bus, thus reflects the Kirchhoff Current Law (KCL). With a given demand $\demand$ this translates to   
\begin{align}
    \nodalgeneration - \sum_\ell \incidence \flow  &=  \demand 
     \resultsin{\lmp} \Forall{n,t}
    \label[constraint]{eq:nodal_balance_lin}
\end{align}
where $\incidence$ is +1 if line $\ell$ starts at bus $n$, -1 if it ends at $n$, 0 otherwise. The nodal generation $\nodalgeneration$ collects the production of all nodal assets, see \cref{eq:nodalgeneration}. The shadow price of the nodal balance constraint mirrors the Locational Marginal Prizes (LMP) $\lmp$ per bus and time step. In a optimal nodal pricing scheme this is the \euro/\megawatthour price which a consumer has to pay.\\

\subsection{Full Lagrangian}
\label{sec:full_lagrangian}
The Lagrangian for the investment model can be condensed to the following expression

\begin{align}
\notag
& \lagrangian\left(\state, \capacity, \lmp, \mu_j \right) = \\ 
\notag
& \qquad + \sum_{i,t} o_i \state + \sum_{i} c_i \capacity  \\
\notag
&\qquad +  \sum_{n,t} \lmp \left( \demand - \nodalgeneration + \sum_\ell \incidence \, \flow \right) \\
&\qquad + \sum_j \mu_j \, h_j \left(\state, \capacity \right)
\label{eq:full_lagrangian}
\end{align}
where $h_j \left(\state, \capacity \right)$ denotes all inequality constraints attached to $\state$ and $\capacity$. In order to impose the Kirchhoff Voltage Law (KVL) for the linearized AC flow, the term 
\begin{align}
    \sum_{\ell,c,t} \cycleprice \, \cycle \, \reactance \, \flow 
\end{align}
can be added to $\lagrangian$, with $\reactance$ denoting the line's impedance and $\cycle$ being 1 if $\ell$ is part of the cycle $c$ and zero otherwise.

The global maximum of the Lagrangian requires stationarity with respect to all variables:
\begin{align}
    \pdv{\lagrangian}{\state} = \pdv{\lagrangian}{\capacity} = 0    
\end{align}

\subsection{Zero Profit Generation}
\label{sec:zero_profit_generation}
For each generator the optimization defines a lower capacity constraint, given by 
\begin{align}
    - \generation &\le 0 \resultsin{\mulowergeneration} \Forall{s,t} 
    \label[constraint]{eq:lower_generation_capacity_constraint}
\end{align} 
\Cref{eq:upper_generation_capacity_constraint,eq:lower_generation_capacity_constraint}, which yield the KKT variables $\muuppergeneration$ and $\mulowergeneration$, imply the complementary slackness,
\begin{align}
\muuppergeneration \left( \generation - \generationpotential \, \capacitygeneration \right)  &= 0  \Forall{s,t} 
\label{eq:complementary_slackness_upper_generation} \\
\mulowergeneration  \, \generation &= 0 \Forall{s,t}
\label{eq:complementary_slackness_lower_generation} 
\end{align}

The stationarity of the generation capacity variable leads to 
\begin{align}
\pdv{\lagrangian}{\capacitygeneration}  = 0 \,\, \rightarrow \,\, 
\capitalpricegeneration =  \sum_t \muuppergeneration \, \generationpotential  \Forall{s}
\label{eq:capex_generation_duality}
\end{align}
and the stationarity of the generation to 
\begin{align}
\pdv{\lagrangian}{\generation} &= 0 \,\, \rightarrow \,\,  
\operationalpricegeneration =  \sum_n \incidencegenerator \, \lmp - \muuppergeneration + \mulowergeneration \Forall{s} \label{eq:opex_duality}
\end{align}

Multiplying both sides of \cref{eq:capex_generation_duality} with $\capacitygeneration$ and using \cref{eq:complementary_slackness_upper_generation} leads to 
\begin{align}
 \capitalpricegeneration \, \capacitygeneration  = \sum_t \muuppergeneration \, \generation \Forall{s} 
 \label{eq:capital_price_generation_sum}
\end{align}
The zero-profit rule for generators is obtained by multiplying \cref{eq:opex_duality} with $\generation$ and using \cref{eq:complementary_slackness_lower_generation,eq:capital_price_generation_sum} which results in 
\begin{align}
  \capitalpricegeneration \, \capacitygeneration + \sum_t \operationalpricegeneration \generation = \sum_{n,t} \lmp \incidencegenerator \generation \Forall{s}
\end{align}
It states that over the whole time span, all OPEX and CAPEX for generator $s$ (left hand side) are payed back by its revenue (right hand side).

\subsection{Zero Profit Transmission System}
\label{sec:zero_profit_flow}

The yielding KKT variables $\muupperflow$ and $\mulowerflow$ are only non-zero if $\flow$ is limited by the transmission capacity in positive or negative direction, i.e. \cref{eq:upper_flow_capacity_constraint} or \cref{eq:lower_flow_capacity_constraint} are binding. For flows below the thermal limit, the complementary slackness 
\begin{align}
\muupperflow \left( \flow - \capacityflow \right)  &= 0 \Forall{\ell,t}
\label{eq:complementary_slackness_upper_flow} \\
\mulowerflow \left( \flow - \capacityflow \right) &=  0 \Forall{\ell,t}
\label{eq:complementary_slackness_lower_flow} 
\end{align}
sets the respective KKT to zero. 

The stationarity of the transmission capacity leads to
\begin{align}
\pdv{\lagrangian}{\capacityflow} = 0 \,\, \rightarrow \,\, 
\capitalpriceflow =  \sum_t \left( \muupperflow - \mulowerflow \right) \Forall{\ell}
\label{eq:capacity_flow_duality}
\end{align}
and the stationarity with respect to the flow to
\begin{align}
    0 &= \pdv{\lagrangian}{\flow}  \\ 
    0 &= - \sum_n \incidence \lmp  + \sum_c \cycleprice \cycle \reactance  - \muupperflow + \mulowerflow \Forall{\ell, t} \label{eq:flow_duality}
\end{align}

When multiplying \cref{eq:capacity_flow_duality} with $\capacityflow$ and using the complementary slackness \cref{eq:complementary_slackness_upper_flow,eq:complementary_slackness_lower_flow} we obtain 
\begin{align}
 \capitalpriceflow \, \capacityflow = \sum_t \left( \muupperflow - \mulowerflow \right)  \, \flow \Forall{\ell} 
 \label{eq:capital_price_flow_sum}
\end{align}
Again we can use this to formulate the zero-profit rule for transmission lines. We multiply \cref{eq:flow_duality} with $\flow$, which finally leads us to 
\begin{align}
\capitalpriceflow \, \capacityflow = - \sum_n \incidence\, \lmp\, \flow + \sum_c \cycleprice\, \cycle\, \reactance\, \flow 
\Forall{\ell} \end{align}
It states that the congestion revenue of a line (first term right hand side) reduced by the cost for cycle constraint exactly matches its CAPEX.

\subsection{Zero Profit Storage Units}
\label{sec:zero_profit_storage_units}

For an simplified storage model, the upper capacity $\capacitystorage$ limits the discharging dispatch $\storagedispatch$, the storing power $\storagecharge$ and state of charge $\storagesoc$ of a storage unit $r$ by 
\begin{align}
    \storagedispatch - \capacitystorage &\le 0  \resultsin{\muupperstoragedispatch} \Forall{r,t} \\
    \storagecharge - \capacitystorage &\le 0  \resultsin{\muupperstoragecharge} \Forall{r,t} \\
    \storagesoc - h_r \, \capacitystorage &\le 0  \resultsin{\muupperstoragesoc} \Forall{r,t}
\end{align}
where we assume a fixed ratio between dispatch and storage capacity of $h_r$. 
The state of charge must be consistent throughout every time step according to what is dispatched and stored, 
\begin{align}
    \notag
    \storagesoc - \efficiencysoc \storageprevioussoc - \efficiencycharge \storagecharge &+ (\efficiencydispatch)^{-1} \storagedispatch = 0 \\
    &\resultsin{\mustateofcharge} \Forall{r,t}
\end{align}

We use the result of Appendix B.3 in \cite{brown_decreasing_2020} which shows that a storage recovers its capital (and operational) costs from aligning dispatch and charging to the LMP, thus 
\begin{align}
    \notag
    \sum_t \operationalpricestorage \, \storagedispatch + \capitalpricestorage \, \capacitystorage = \sum_t \lmp \incidencestorage &\left(\storagedispatch - \storagecharge \right) \Forall{r,t}
\end{align}
The stationarity of the dispatched power leads us to  
\begin{align}
    \notag
    \pdv{\lagrangian}{\storagedispatch} &= 0 \\
    \operationalpricestorage - \sum_n \lmp \, \incidencestorage - \mulowerstoragedispatch + \muupperstoragedispatch + (\efficiencydispatch )^{-1} \mustateofcharge &= 0 \;  \forall r,t
    \label{eq:stationarity_storagedispatch}
\end{align}
which we  can use to define the revenue which recovers the CAPEX at $r$, 
\begin{align}
    \notag
    \capitalpricestorage \, \capacitystorage = \sum_t \left(\muupperstoragedispatch - \mulowerstoragedispatch  + (\efficiencydispatch )^{-1} \mustateofcharge \right) \storagedispatch \\
    - \sum_t \lmp \incidencestorage  \storagecharge \Forall{r} 
    \label{eq:no_profit_capex_storage2}
\end{align}


When applying the cost allocation scheme \cref{eq:general_scheme}, it stands to reason to assume that when a storage charges power, it does not supply any demand. Thus consumers only pay storage units in times the storage dispatches power. Hence, we restrict the allocatable revenue per storage unit to the first term in \cref{eq:no_profit_capex_storage,eq:no_profit_capex_storage2}. This allocates then the CAPEX of $r$ plus the costs $\remainingcost^E_r$ it needs to buy the charging power, 
\begin{align}
        \capexstorage_r + \remainingcost^E_r = \sum_t \left(\muupperstoragedispatch - \mulowerstoragedispatch  + (\efficiencydispatch )^{-1} \mustateofcharge \right) \storagedispatch 
\end{align} 
In charging times the total of remaining costs $\remainingcost^E_r$ is spent to power from other assets. These costs scale with the amount of installed storage capacity.
Note that it would be possible to incorporate this redistribution into the cost allocation, by replacing the demand $\demand$ with the power charge $\storagecharge$ in \cref{eq:general_scheme}. Then, the derived payments that a storage unit $r$ has to pay to asset $i$ is given by $\cost_{r \rightarrow i}$. The sum of those payments due to $r$ will the sum up to $\remainingcost^E_r$.

\subsection{Proof: Equivalence of local and imported prices}
\label{sec:proof_equivalence}

We start with \cref{eq:flow_duality} which we recall here, 
\begin{align}
    \notag
    0 = &- \sum_m \incidence[m] \lmp[m]  + \sum_c  \cycleprice \cycle \reactance  \\
    &- \muupperflow + \mulowerflow \Forall{\ell,t} 
\end{align}
It states that the price difference between two adjacent buses minus the price for the KVL, is the revenue per line $\ell$, $\left(- \muupperflow + \mulowerflow\right)$. We multiply the equation by the flow allocation $\allocateflow$ and obtain 
\begin{align}
    \notag
    0 =& - \allocateflow \sum_m \incidence[m] \lmp[m]  \\
    \notag
    &+ \allocateflow \sum_c  \cycleprice \cycle \reactance  \\
    &- \allocateflow \left(\muupperflow - \mulowerflow\right) \Forall{\ell,t} 
    \label{eq:flow_duality2}
\end{align}  
The allocation $\allocateflow$ defined in \cref{eq:allocate_flow}follows the linear power flow laws. We slightly reformulate the expression to
\begin{align}
    \allocateflow = \sum_{m'} \ptdf[m'] \left( \allocatepeer[m' \rightarrow n]  - \delta_{n m'} \demand \right) \Forall{\ell, n, t}
    \label{eq:allocate_flow2}
\end{align}
and insert it into \cref{eq:flow_duality2}. When taking the sum over all lines $L$, the first term yields  
\begin{align}
    \notag
    &- \sum_{\ell, m'} \ptdf[m'] \left( \allocatepeer[m' \rightarrow n]  - \delta_{n m'} \demand \right) \sum_m \incidence[m] \lmp[m] \\
    \notag
    =& - \sum_{\ell, m', m} \ptdf[m'] \left( \allocatepeer[m' \rightarrow n]  - \delta_{n m'} \demand \right) \incidence[m] \lmp[m] \\
    \notag
    =& - \sum_{m', m} \delta_{m m'} \left( \allocatepeer[m' \rightarrow n]  - \delta_{n m'} \demand \right) \lmp[m] \\
    \notag
    =& - \sum_{m} \left( \allocatepeer - \delta_{n m} \demand \right) \lmp[m] \\
    =& - \sum_{m} \allocatepeer \lmp[m] + \demand \lmp 
\end{align}
where in the third step we used the relation $\sum_\ell \ptdf \incidence[m] = \delta_{n m}$.
The second term in \cref{eq:flow_duality2} vanishes as the basis cycles $\cycle$ are the kernel of the PTDF, $\sum_{\ell} \cycle \ptdf  = 0 \; \forall c,n$. Thus, we end up with 
\begin{align}
    \notag
    \demand \lmp =& \sum_{m} \allocatepeer \lmp[m] \\
    +& \sum_\ell \allocateflow \left(\muupperflow - \mulowerflow\right) \Forall{n,t}
\end{align}
This relation shows that for any P2P allocations $\allocatepeer$ the combined price of the imported power is always be the same as the locational price. The representation matches the findings in \cite{wu_locational_2005}. However, the latter builds its formulation on a evenly distributed slack, which translated to a peer-to-peer allocation $\allocatepeer$ corresponding to the non-local Equivalent Bilateral Exchanges \cite{galiana_transmission_2003}. However, the representation here holds only true any $\allocatepeer$ if the corresponding flow allocation $\allocateflow$ follow the power flow laws, \ie are defined as in \cref{eq:allocate_flow,eq:allocate_flow2}. 

Naturally, the power production of the supplying node $m$ decomposes into contributions of assets, following the definition in \cref{eq:allocate_production}. At the same time, the LMP at $m$ decomposes into asset related prices (\cref{eq:opex_duality,eq:stationarity_storagedispatch}). This finally reproduces the allocations of \cref{tab:cost_allocation_map} and results in 
\begin{align}
    \demand \, \lmp =  \sum_{\circ, i} \allocatecost 
\end{align}

\section{Power Allocation}
\label{sec:net_ap}

\newcommand{\incidenceM}{K}
\newcommand{\flowM}{f}
\newcommand{\injectionM}{p}
\newcommand{\slackM}{k}
\newcommand{\DirectedIncidence}{\mathcal{K}}
\newcommand{\InverseAPInjection}{\mathcal{J}}
\newcommand\diag[1]{\operatorname{diag}\left(#1\right)}

Allocating net injections using the AP method is derived from \cite{achayuthakan_electricity_2010}. In a lossless network the downstream and upstream formulations result in the same P2P allocation which is why we restrict ourselves to the downstream formulation only. In a first step we define a time-dependent auxiliary matrix $\InverseAPInjection_t$ which is the inverse of the $N\times N$ with directed power flow $m \rightarrow n$ at entry $(m, n)$ for $m \ne n$ and the total flow passing node $m$ at entry $\left( m, m\right)$ at time step $t$. Mathematically this translates to

\begin{align}
\InverseAPInjection_t = \left( \diag{\injectionM^+} + \DirectedIncidence^- \diag{\flowM} \, \incidenceM \right)_t^{-1} 
\end{align}
where $\DirectedIncidence^-$ is the negative part of the directed Incidence matrix $\DirectedIncidence_{n,\ell} = \text{sign}\left( f_\ell \right)  \incidence$. Then the P2P allocation for time step $t$ is given by
\begin{align}
\allocatepeer = \InverseAPInjection_{m,n,t} \, \netproduction[m] \, \netconsumption
\end{align}

\section{Working Example}
The following figures contain more detailed information about the peer-to-peer cost allocation discussed in \cref{sec:application_case}. The cost or prices payed by consumers are indicated by the region color. The dedicated revenue is displayed in proportion to the size of cycles (for assets attached to buses) or to the thickness of transmission branches.    
\begin{figure*}
    \centering
    \begin{subfigure}[c]{.49\linewidth}
        \includegraphics[width=\linewidth]{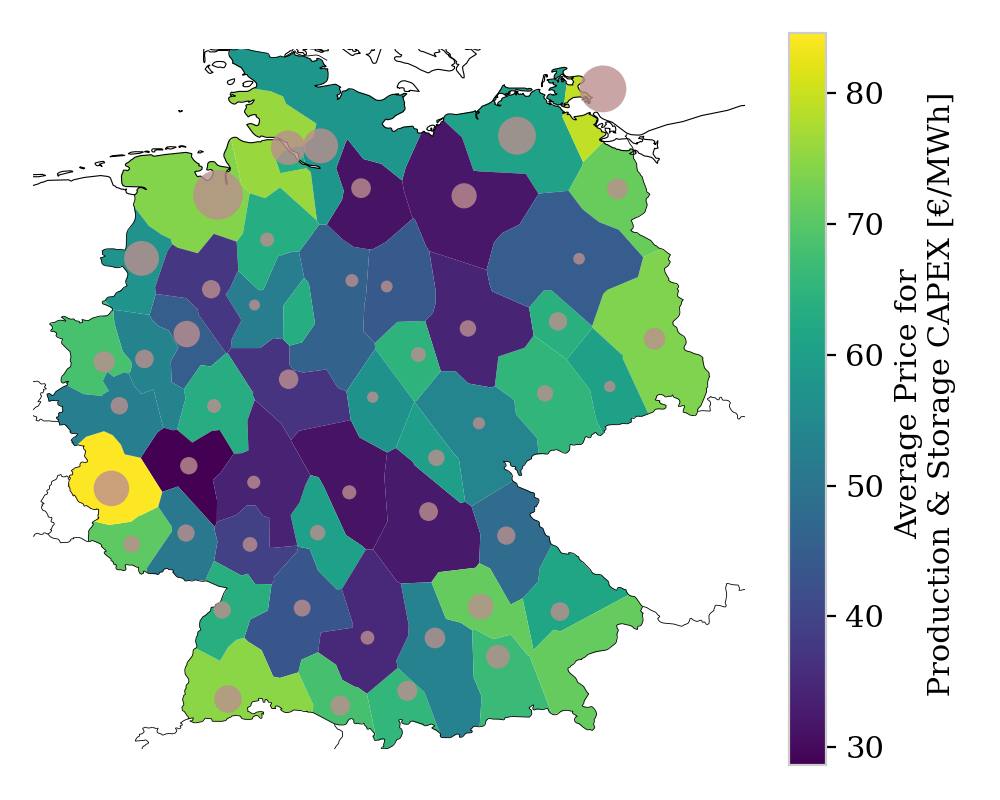}
        \subcaption{All production and storage technologies}
        \label{fig:total_capex}
    \end{subfigure}
    \begin{subfigure}[c]{.49\linewidth}
        \includegraphics[width=\linewidth]{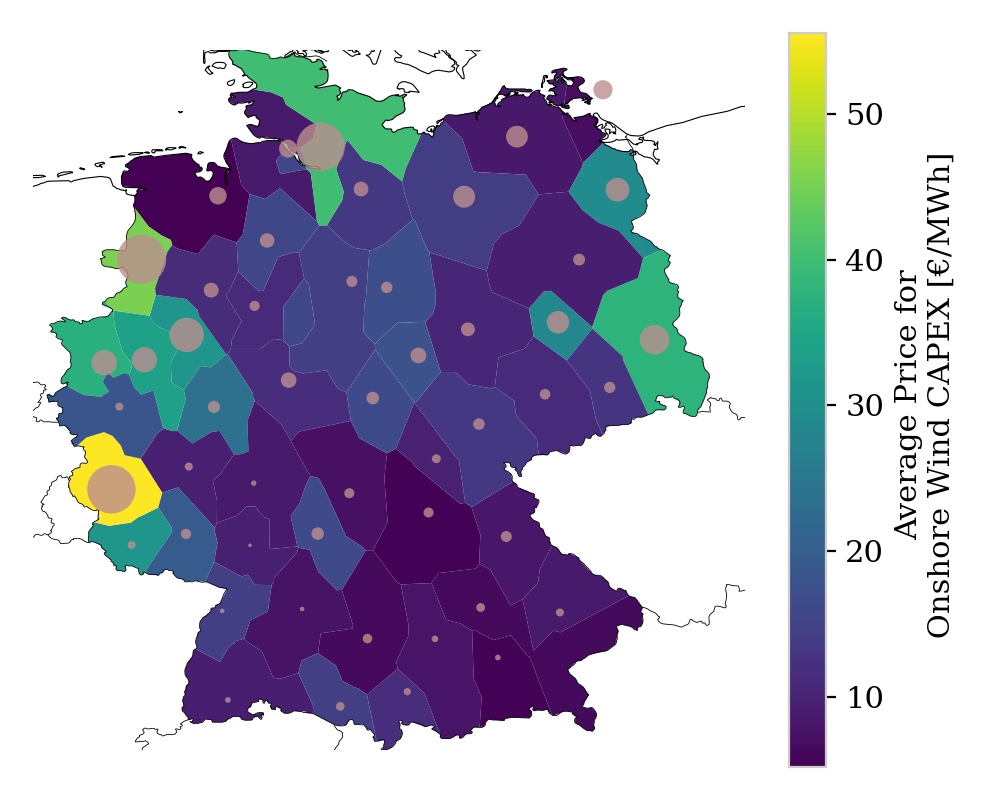}
        \subcaption{Onshore Wind}
        \label{fig:onshore_capex}
    \end{subfigure}
    \begin{subfigure}[c]{.49\linewidth}
        \includegraphics[width=\linewidth]{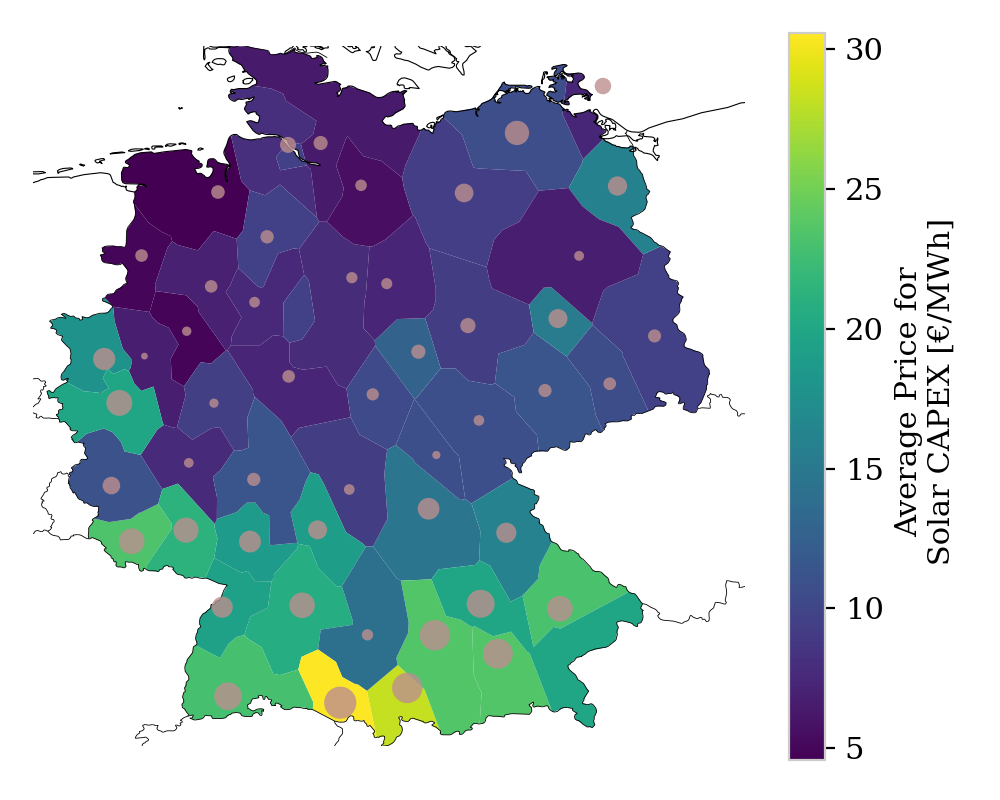}
        \subcaption{Solar}
        \label{fig:solar_capex}
    \end{subfigure}
    \begin{subfigure}[c]{.49\linewidth}
        \includegraphics[width=\linewidth]{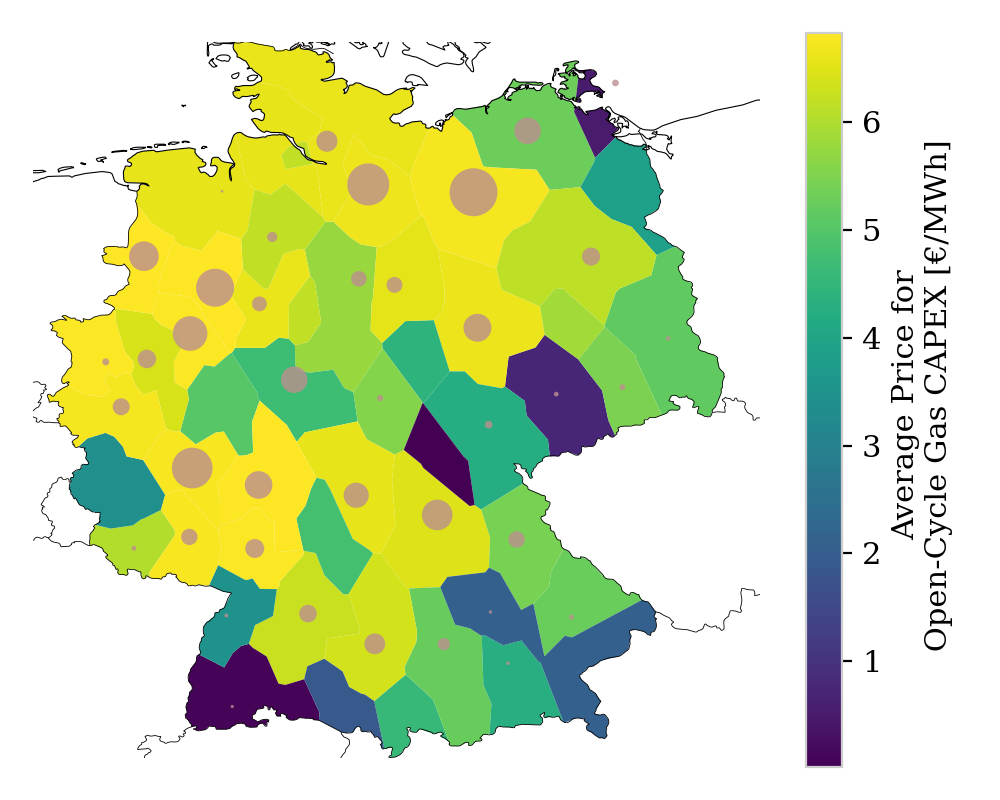}
        \subcaption{OCGT}
        \label{fig:ocgt_capex}
    \end{subfigure}
    \caption{Average \textbf{CAPEX allocation} per MWh, $\sum_t \allocatecapex / \sum_t \demand$  for all production and storage assets (a), onshore wind (b), solar (c) and OCGT (d). Average allocated CAPEX per MWh within the regions are indicated by the color, the revenue per production asset is given by the size of the circles at the corresponding bus.}
    \label{fig:capex_price}
\end{figure*}

\begin{table*}[h]
    \centering
    \begin{tabular}{lllr}
\toprule
     &    & o [\euro/MWh] &  c [k\,\euro/MW]$^*$ \\
{} & carrier &               &                      \\
\midrule
Generator & Open-Cycle Gas &       120.718 &               47.235 \\
     & Offshore Wind (AC) &         0.015 &              204.689 \\
     & Offshore Wind (DC) &         0.015 &              230.532 \\
     & Onshore Wind &         0.015 &              109.296 \\
     & Run of river &               &              270.941 \\
     & Solar &          0.01 &               55.064 \\
Storage & Hydrogen Storage &               &              224.739 \\
     & Pumped Hydro &               &              160.627 \\
     & Battery Storage &               &              133.775 \\
Line & AC &               &                0.038 \\
     & DC &               &                0.070 \\
\bottomrule
\end{tabular}
    
    \caption{Operational and capital price assumptions for all type of assets used in the working example. The capital price for transmission lines are given in [k\,\euro/MW/km]. The cost assumptions are retrieved from the PyPSA-EUR model \cite{horsch_jonas_pypsa-eur_2020}.}
    \label{tab:cost_assumptions}
\end{table*} 

\begin{figure}
    \includegraphics[width=\linewidth]{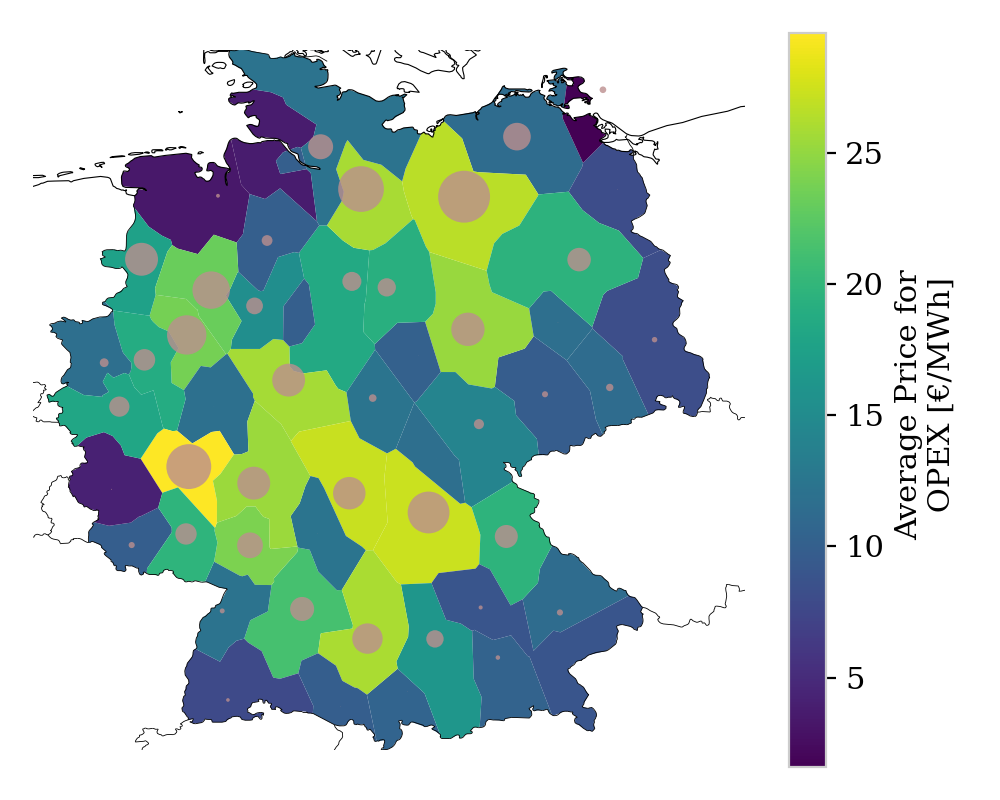}
    \caption{Average \textbf{OPEX allocation} per consumed MWh, $\sum_t \allocateopex/\sum_t \demand$. The effective prices for OPEX are indicated by the color of the region, the circles are drawn in proportion to the revenue per regional generators and storages. As OCGT is the only allowed fossil based technology, the drawn allocation is proportional to OPEX allocation of OCGT generators.}
    \label{fig:opex_price}
\end{figure}

\begin{figure}
    \includegraphics[width=\linewidth]{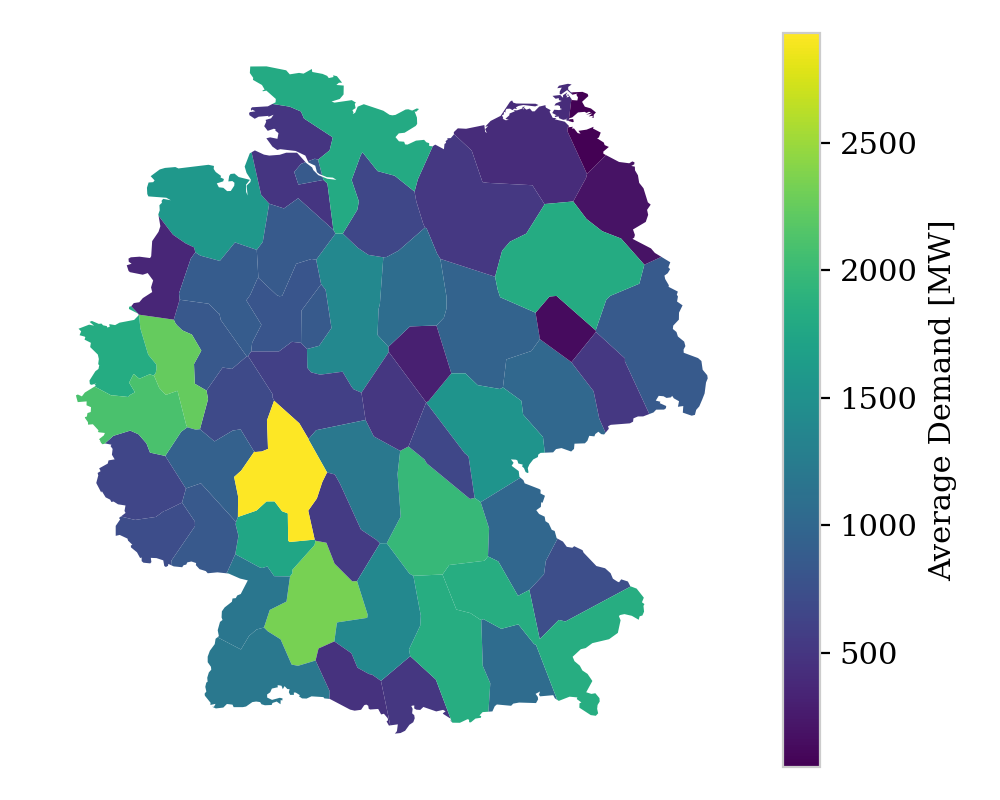}
    \caption{Average demand, $\sum_t \demand/T$ per regions. The regions with high population densities and larger areas reveal a higher demand.}
    \label{fig:average_demand}
\end{figure}

\begin{figure}
    \includegraphics[width=\linewidth]{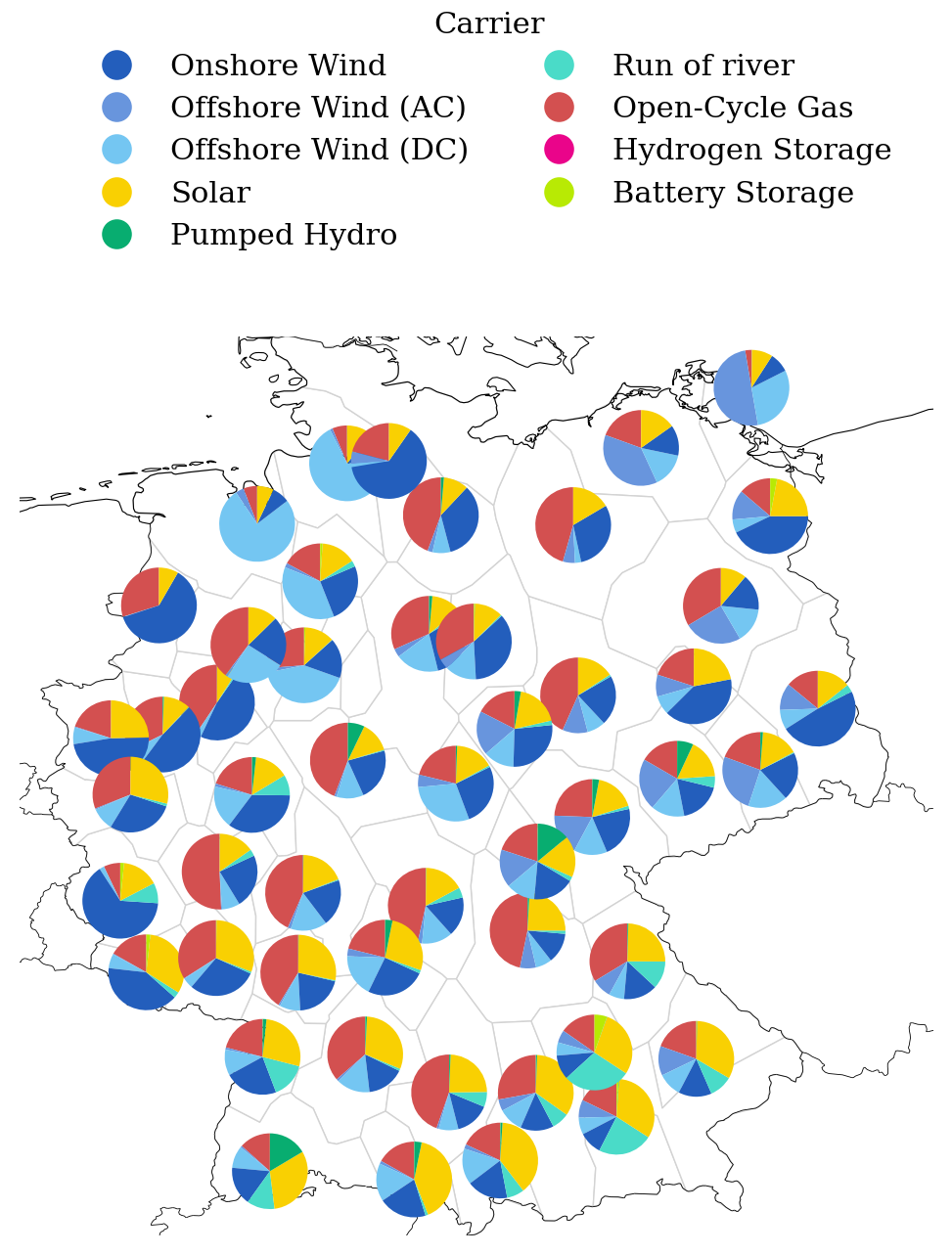}
    \caption{Average power mix per region calculated by Average Participation. Coastal regions are mainly supplied by local offshore and onshore wind farms. Their strong power injections additionally penetrate the network up to the southern border. In the middle and South, the supply is dominated by a combination of OCGT and solar power.}
    \label{fig:power_mix}
\end{figure}

\begin{figure}
    \vspace{2cm}
    \includegraphics[width=\linewidth]{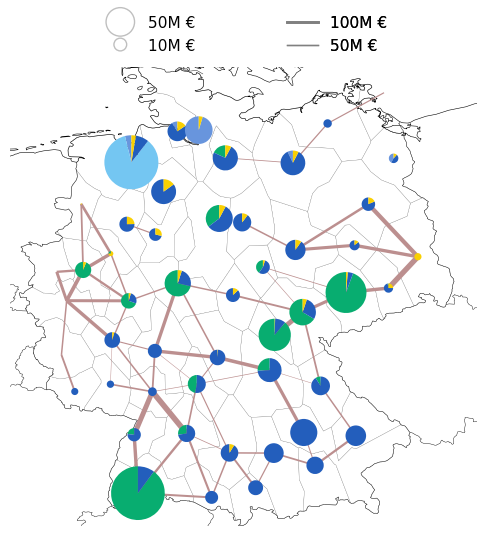}
    \caption{Total costs for subsidy $\subsidycost$ resulting from lower capacity expansion bounds (brownfield constraints). The figure shows the built infrastructure that does not gain back its CAPEX from its market revenue, but is only built due to lower capacity limits.}
    \label{fig:subsidy}
\end{figure}

\begin{figure*}
    \includegraphics[width=\linewidth]{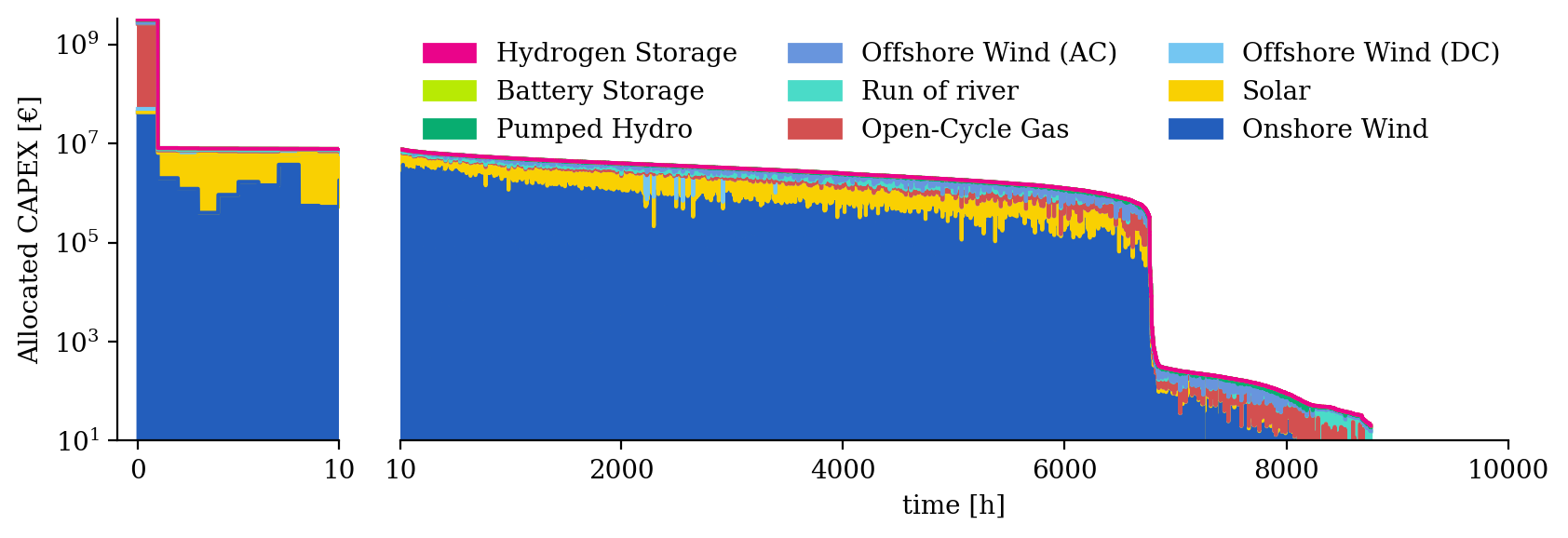}
    \caption{Duration curve of the CAPEX allocation for production and storage technologies. Hours are sorted by the total amount of allocated expenditure. With 2.7 bn~\euro\, the first value pushes investments extraordinarily high. Due to low renewable potentials, it is dominated by CAPEX for OCGT which receives 92\% of the payments. This hour alone occasions about the half of all OCGT CAPEX. \Cref{fig:capex_duration_curve} gives a detailed picture of the operational state at this time-step. The following 7000 time-steps are dominated by revenues for onshore wind and reveal a rather even distribution. In hours of low CAPEX allocation (after the second drop) spending for OCGT start to increase again. These time-steps however play a minor role.}
    \label{fig:capex_duration_curve}
\end{figure*}

\begin{figure}
    \includegraphics[width=\linewidth]{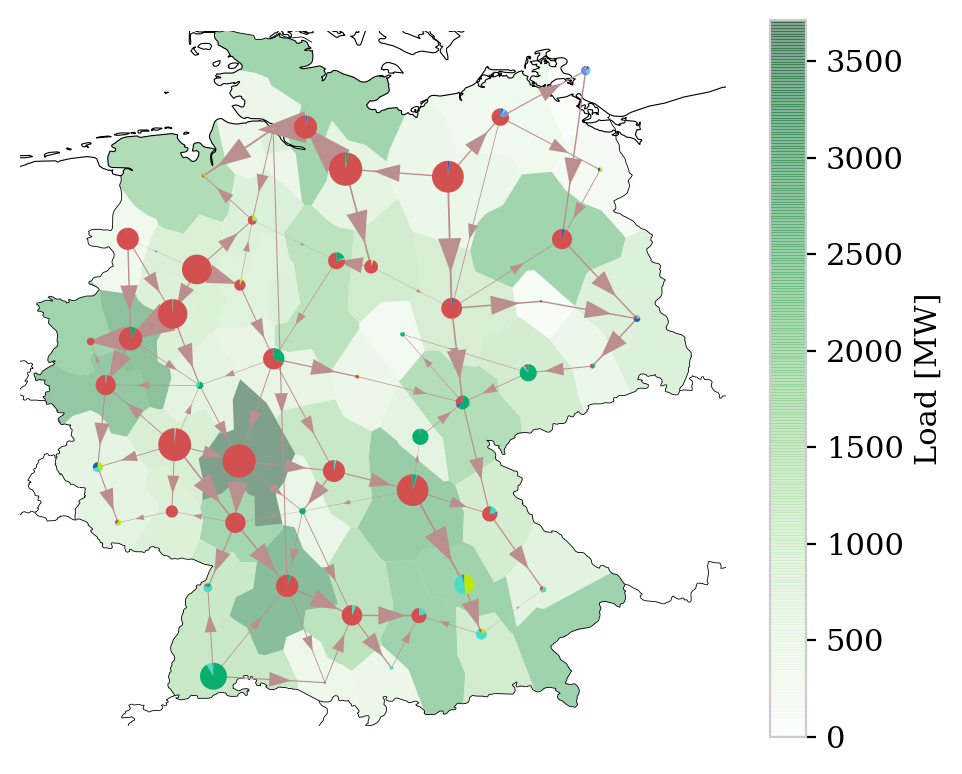}
    \caption{Production, flow and consumption in the system at the hour with the highest allocated expenditures. The size of the circles are proportional to the power production at a node. Size of arrows are proportional to the flow on the transmission line. The depicted hour corresponds to the first value in the duration curve in \cref{fig:capex_duration_curve}.}
    \label{fig:operation_high_expenditure}
\end{figure}

\begin{figure}
    \includegraphics[width=\linewidth]{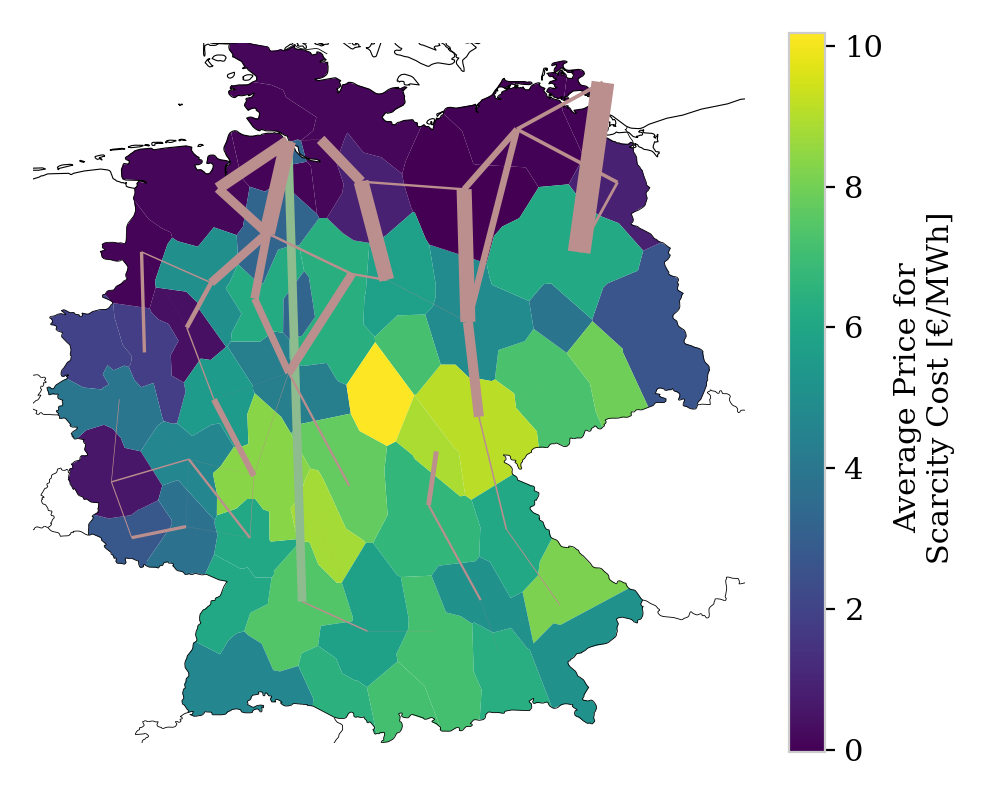}
    \caption{Average \textbf{allocated transmission scarcity cost} per consumed MWh, $\sum_t \scarcitycost_{n \rightarrow \ell, t} / \sum_t \demand $. This scarcity cost results from the upper transmission expansion limit of 25\%. The costs are indicated by the regional color.  The lines are drawn in proportion to revenue dedicated to scarcity cost. }
    \label{fig:branch_scarcity_price}
\end{figure}

\begin{figure*}
    \centering
    \begin{subfigure}[c]{.49\linewidth}
        \includegraphics[width=\linewidth]{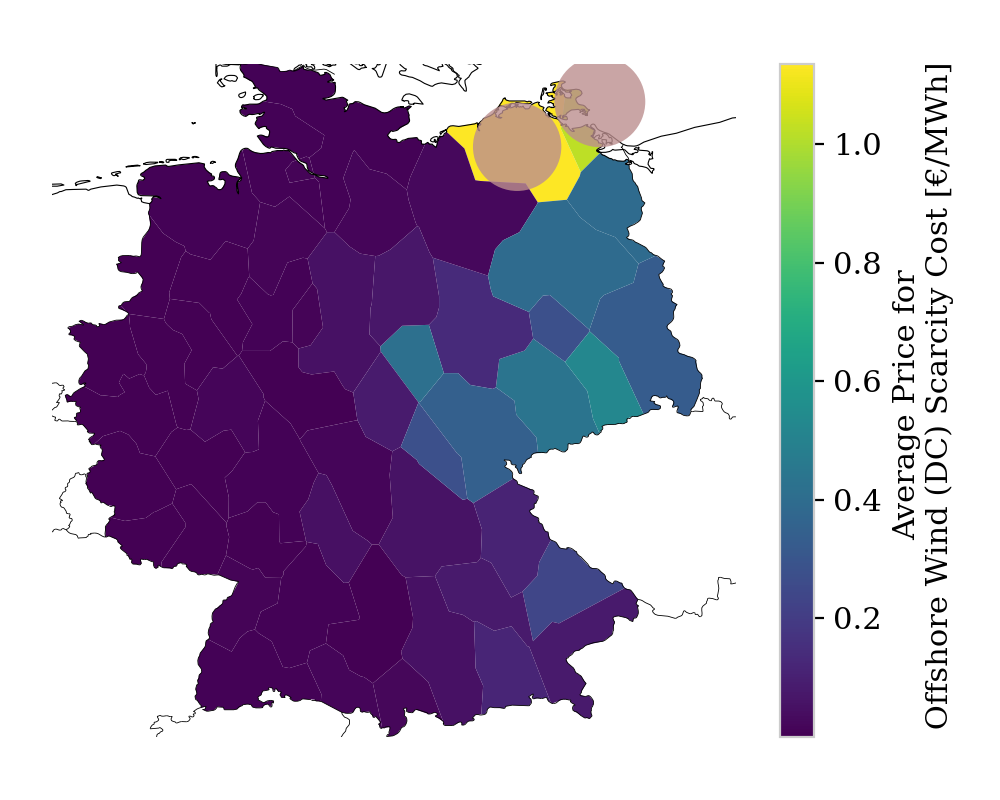}
        \subcaption{Offshore Wind}
        \label{fig:offwind-dc_generator_scarcity_cost}
    \end{subfigure}
    \begin{subfigure}[c]{.49\linewidth}
        \includegraphics[width=\linewidth]{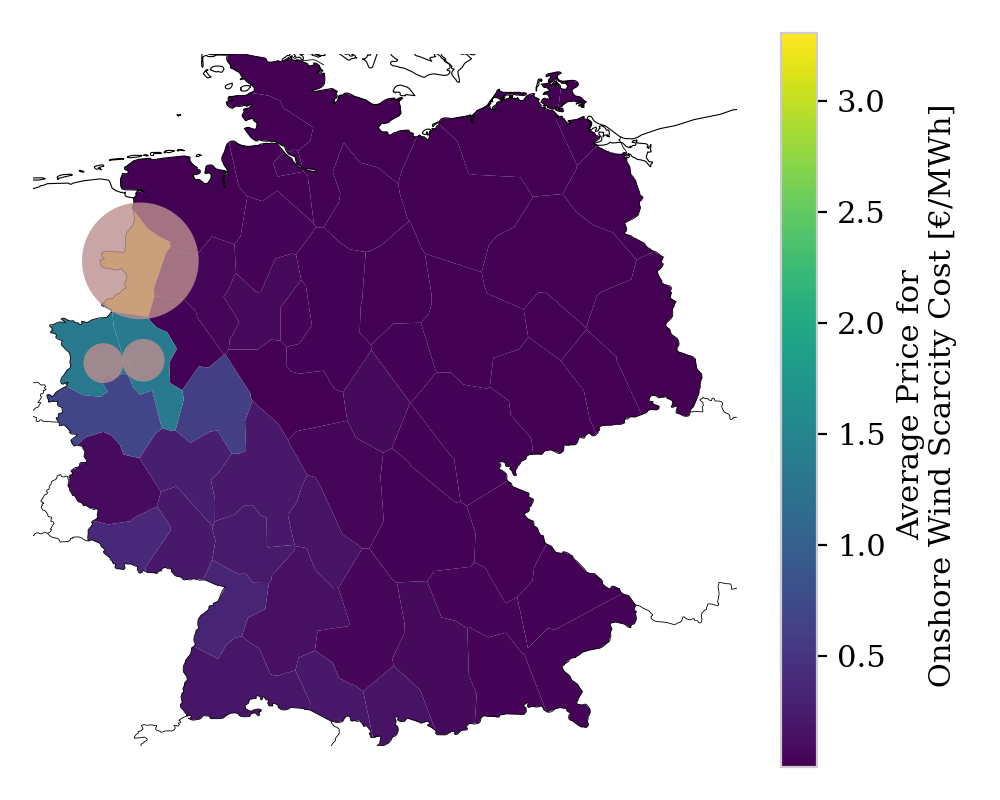}
        \subcaption{Onshore Wind}
        \label{fig:onwind_generator_scarcity_cost}
    \end{subfigure}
    \begin{subfigure}[c]{.49\linewidth}
        \includegraphics[width=\linewidth]{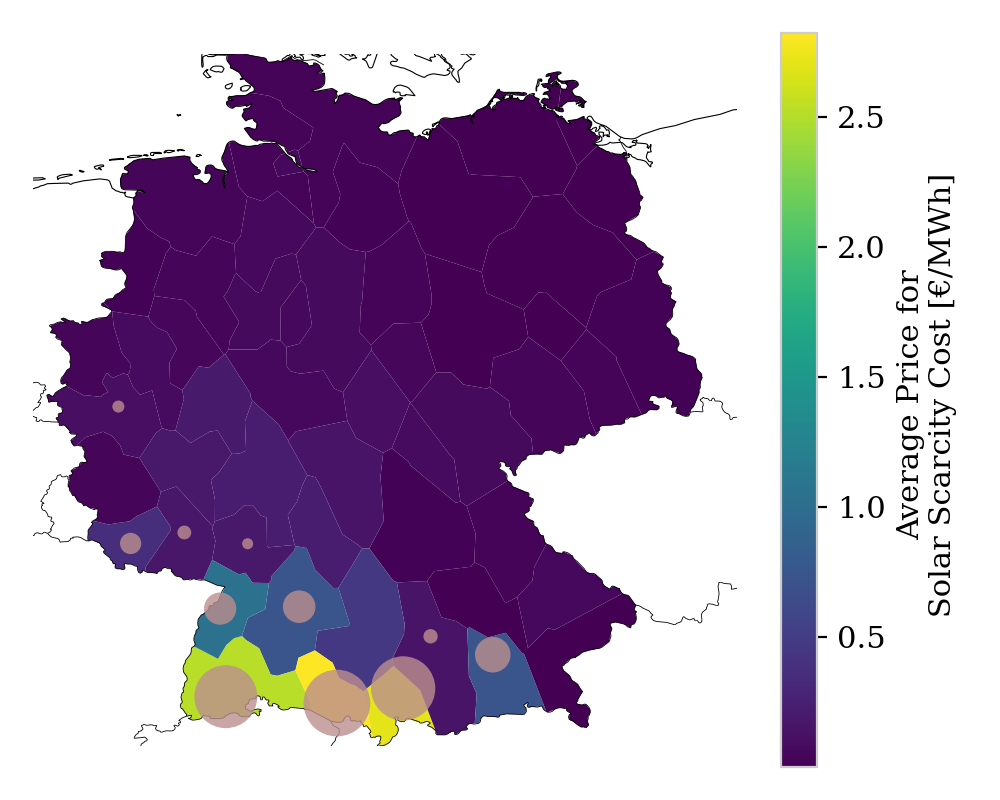}
        \subcaption{Solar}
        \label{fig:solar_generator_scarcity_cost}
    \end{subfigure}
    \begin{subfigure}[c]{.49\linewidth}
        \includegraphics[width=\linewidth]{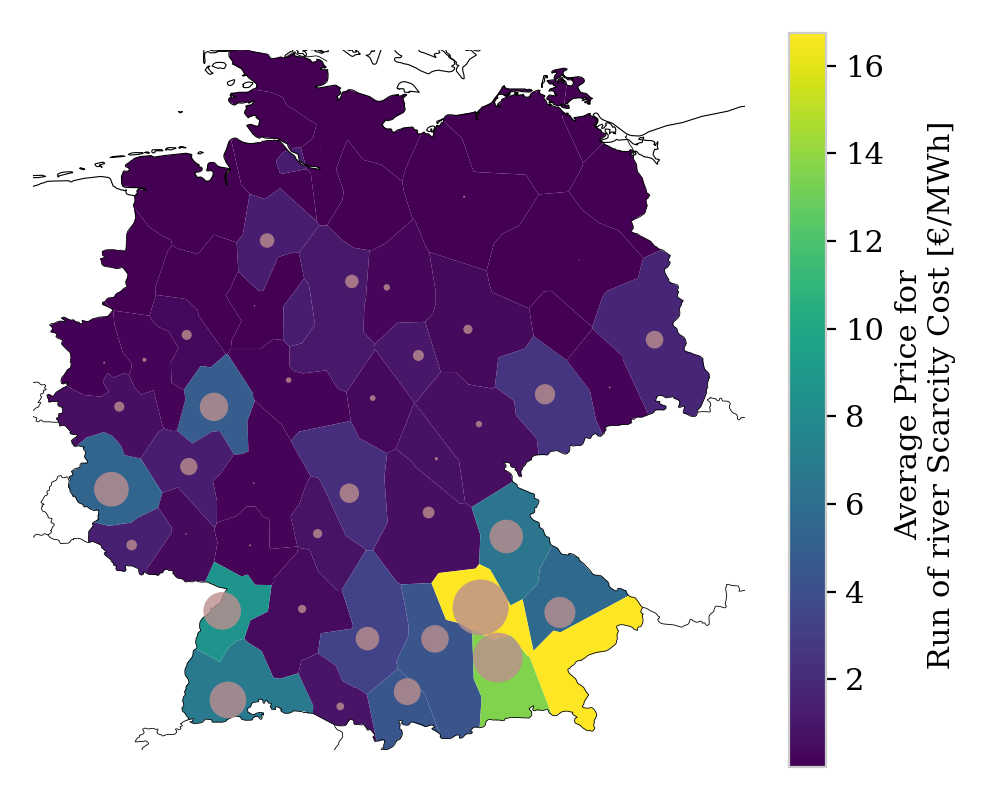}
        \subcaption{Run-of-River}
        \label{fig:ror_generator_scarcity_cost}
    \end{subfigure}
    \caption{Average \textbf{allocated scarcity cost} per consumed MWh, $\sum_t \allocatescarcitycost / \sum_t \demand$. These cost result from land use restrictions for offshore wind, onshore wind,  solar, run-of-river. The cost per MWh are indicated by the color of a region. The revenue per production asset is given by the size of the circle at the corresponding bus.}
    \label{fig:scarcity_price}
\end{figure*}

\clearpage
\printbibliography

\end{document}